\documentclass[sigconf]{acmart}

\usepackage{enumitem}
\usepackage{natbib}  
\usepackage{hyperref}
\usepackage{graphicx}
\usepackage{amsmath}
\usepackage{multirow}
\usepackage{makecell}
\usepackage{subcaption}
\usepackage{float}
\usepackage{xcolor}

\setlist[itemize]{leftmargin=5.5mm}
\setlist[enumerate]{leftmargin=5.5mm}

\AtBeginDocument{%
  }

\copyrightyear{2026}
\acmYear{2026}
\setcopyright{cc}
\setcctype{by}
\acmConference[CHI '26]{Proceedings of the 2026 CHI Conference on Human Factors in Computing Systems}{April 13--17, 2026}{Barcelona, Spain}
\acmBooktitle{Proceedings of the 2026 CHI Conference on Human Factors in Computing Systems (CHI '26), April 13--17, 2026, Barcelona, Spain}
\acmPrice{}
\acmDOI{10.1145/3772318.3791056}
\acmISBN{979-8-4007-2278-3/2026/04}

\begin{document}

 \newcommand{\out}[1]{#1} 

\newcommand{\papertitle}{An Empirical Study to Understand How Students Use ChatGPT for Writing Essays}



\newcommand{\rqone}{How do students use ChatGPT in essay writing?}
\newcommand{\rqtwo}{What factors can predict how students use ChatGPT?}
\newcommand{\rqthree}{How does students' ChatGPT usage manifest in their writing?}
\newcommand{\rqfour}{How does students' ChatGPT usage shape their perception of the writing experience?}

\newcommand{\datalink}{https://doi.org/10.7910/DVN/K6PSHK}

\newcommand{\datalinkanon}{[LINK REDACTED FOR ANONYMITY]}
\title{\papertitle}

\author{Andrew Jelson}
\email{jelson9854@vt.edu}
\affiliation{%
  \institution{Virginia Tech}
  \city{Blacksburg}
  \state{Virginia}
  \country{USA}
  \postcode{24061}
}

\author{Daniel Manesh}
\email{danielmanesh@vt.edu}
\affiliation{%
  \institution{Virginia Tech}
  \city{Blacksburg}
  \state{Virginia}
  \country{USA}
  \postcode{24061}
}

\author{Alice Jang}
\email{ajjang@vt.edu}
\affiliation{%
  \institution{Virginia Tech}
  \city{Blacksburg}
  \state{Virginia}
  \country{USA}
  \postcode{24061}
}

\author{Daniel Dunlap}
\email{dunlapd@vt.edu}
\affiliation{%
  \institution{Virginia Tech}
  \city{Blacksburg}
  \state{Virginia}
  \country{USA}
  \postcode{24061}
}

\author{Young-Ho Kim}
\email{yghokim@younghokim.net}
\affiliation{%
  \institution{NAVER AI Lab}
  \city{Seongnam}
  \country{South Korea}
}

\author{Sang Won Lee}
\authornote{Sang Won Lee conducted this work while at NAVER AI Lab as a visiting scholar.}
\email{sangwonlee@vt.edu}
\affiliation{%
  \institution{Virginia Tech}
  \city{Blacksburg}
  \state{Virginia}
  \country{USA}
  \postcode{24061}
}

\renewcommand{\shortauthors}{Jelson, Manesh, Lee}



\begin{abstract}
As large language models (LLMs) become widespread, students increasingly turn to systems like ChatGPT for writing tasks. Educators worry that this reliance may reduce critical engagement with writing and hinder students' learning processes. Although datasets exist on students’ use of LLMs for writing, how they functionally use ChatGPT in detail---and how this usage shapes their writing and perceptions---remains underexplored. We conducted an online study (n=77) in which students wrote an essay using an in-house ChatGPT we developed to capture their queries. Through qualitative analysis, we identified the types of assistance students sought and presented patterns of use, ranging from asking for opinions on a topic to delegating the entire writing task to ChatGPT. We also found that students' writing self-efficacy influenced their querying patterns and that levels of ownership and creativity varied depending on how they used ChatGPT. This study contributes empirical data to ongoing discussions about how writing education should incorporate or regulate LLM-powered tools.

\end{abstract}




\maketitle

\section{Introduction}

Writing is fundamental to effective learning, allowing learners to critically engage with the topics they study~\cite{emig1977writing, Applebee1984Writing}, with the pedagogical value being the process itself, planning arguments, translating ideas into words, and revising to clarify concepts~\cite{flower_cognitive_1981}. The emergence of generative AI (GenAI) tools, such as ChatGPT, Google Gemini, and Claude, has disrupted traditional educational paradigms by allowing students to offload writing tasks to AI~\cite{herman2022end,marche2022college}. Educators have expressed growing concerns about how students might use GenAI when instructors give students writing assignments (e.g. reflective essays) designed to facilitate students' critical engagement with a topic~\cite{AlAfnan2023ChatGPT, Perkins2023Academic, Sallam2023ChatGPT}.                         

\begin{figure*}[ht!]
    \centering
    \includegraphics[width=1.025\linewidth]{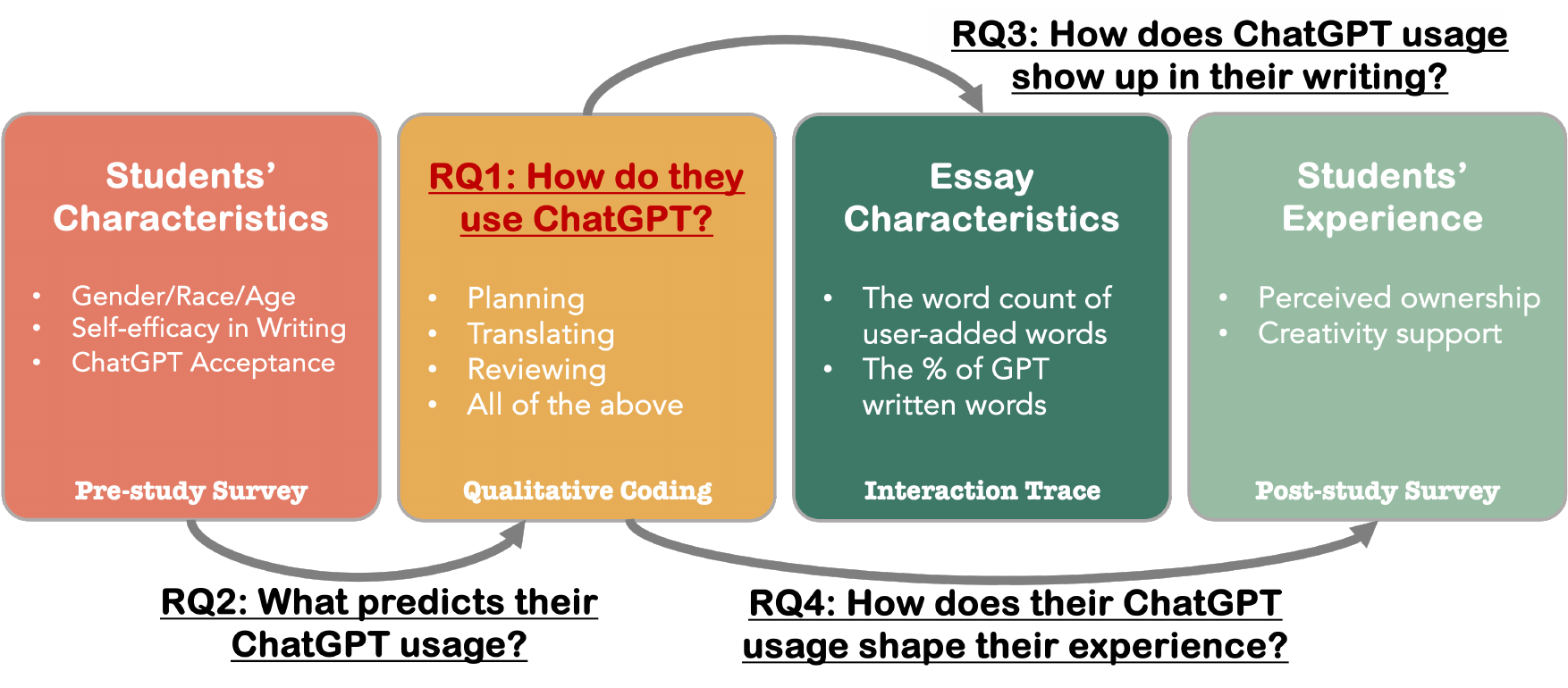}
    \caption{The overview of Research Questions}
    \Description{Conceptual framework showing relationships between four research components. Arrows indicate correlational relationships between usage patterns, writing behaviors, and outcomes. Boxes show the different data collected: Student Characteristics, Qualitative Coding, Essay Characteristics, and Student Experience}
    \label{fig:rqs}
\end{figure*}

Understanding the potential risks and benefits of using LLM-powered tools is challenging because student usage remains understudied. As a result, educators' ability to assess the impact of LLMs on student learning is limited. 
While some uses of ChatGPT may be less problematic than generating an entire essay (e.g., spelling correction), even limited usage can still negatively impact the learning process, depending on the learning context. 
For example, asking an LLM to choose one perspective on a divisive topic can deprive students of the opportunity to think critically about opposing viewpoints. 
However, these expectations remain speculative and are not necessarily evidence-based~\cite{wang_effect_2025, shibani_ai-assisted_2024}.
Therefore, in order for educators to assess its impact on learning, it is crucial to understand how students use ChatGPT based on their objectives and usage patterns.
While students’ actual use of GenAI for writing has been studied~\cite{goldi_support_2024, wilbers_overall_2024, chen_coachgpt_2025} and made available through datasets~\cite{han_recipe4u_nodate, lee_coauthor_2022, liu_detectability_2024}, a detailed analysis of how they functionally use GenAI, what factors predict their GenAI usage, and how such usage influences their writing and perceptions remains underexplored.

This study addresses these gaps by observing students writing an argumentative essay with trace data collection and integrated analysis. We conducted an online study where 77 college students were asked to write an essay using ChatGPT, which was accessible on a custom online platform that we developed to capture the queries they made to ChatGPT---queries that are typically hidden from instructors. In addition, we captured their keystrokes, copying, and pasting behaviors to follow how the writing process was affected by ChatGPT responses.  This data-driven approach captures usage patterns across the writing process—rather than relying on self-reports—and allows us to connect user characteristics with behavioral patterns and resulting essay characteristics. We address the following research questions (shown in Figure \ref{fig:rqs}).

\begin{itemize}
    \item \textbf{RQ1:} \textbf{\rqone}
\end{itemize}
\noindent We categorized all ChatGPT queries based on \citeauthor{flower_cognitive_1981}'s Cognitive Process Theory of Writing Model ~\cite{flower_cognitive_1981}, which structures the cognitive process into three categories: \textit{Planning}, \textit{Translating}, and \textit{Reviewing}. 
We also added another category, \textit{All}, to account for cases when a student relies on ChatGPT to write the entire essay or a subset of the essay, which delegates all three types of tasks at once (e.g., `\textit{Write an essay in response to the following prompt.}').
This taxonomy reveals the types of writing tasks students delegate to ChatGPT and provides a comprehensive overview of usage patterns across the writing process.
In addition, the publicly available dataset collected in this study will allow educators and researchers to identify GenAI usage that they may find problematic or beneficial for student learning and can enable follow-up studies on how instructors should incorporate GenAI into writing pedagogy.

In addition, 
we examined how individual student characteristics relate to ChatGPT usage patterns.
\begin{itemize}[left=.3cm, labelsep=0.15cm]
    \item \textbf{RQ2:} \textbf{\rqtwo}
\end{itemize}
\noindent In particular, we used two specific constructs to account for individual differences, along with demographic information: self-efficacy in writing (SEWS) and the perceived acceptance of ChatGPT measured by the Technology Acceptance Model (TAM). Writing self-efficacy reflects a student's confidence in their ability to perform various writing tasks. Prior research indicates that students with lower self-efficacy in writing are more inclined to seek external assistance or rely on tools to simplify the writing process~\cite{mccarthy_self-efficacy_nodate, woodrow_college_2011}. TAM captures attitudes towards perceived usefulness and ease of use for a technology, factors that predict actual usage~\cite{saif_chat-gpt_2024}. By combining these surveys with demographic data, 
we investigate how students' attitudes toward GenAI and their writing confidence correlate with usage patterns.

Based on usage patterns, we grouped participants into six groups to answer the following:
\begin{itemize}[left=.3cm, labelsep=0.15cm]
    \item \textbf{RQ3:} \textbf{\rqthree}
\end{itemize}
\noindent 
We analyzed the interaction traces to understand how the essays were composed , examining three characteristics: word count, essay composition (student versus ChatGPT), and readability scores measured with the Flesch-Kincaid Grade Level~\cite{flesch_new_1948} and Dale-Chall~\cite{dale_formula_1948} metrics. By comparing these characteristics across clusters, we discuss how different usage patterns relate to essay length and linguistic complexity.

Lastly, we investigated how ChatGPT use relates to students' subjective writing experience, addressed in the following research question: 
\begin{itemize}[left=.3cm, labelsep=0.15cm]
    \item \textbf{RQ4:} \textbf{\rqfour}
\end{itemize}
\noindent 
We assessed two dimensions of the subjective writing experience: perceived ownership (PO)~\cite{avey_psychological_2009, chantal_psychological_2012}---the students' sense of ownership of the essay---and the Creativity Support Index (CSI)~\cite{cherry_quantifying_2014}---their reflection on how ChatGPT supported their writing practice.
PO is particularly important in educational settings, where feelings of authorship and accountability are closely related to learning outcomes and academic integrity~\cite{chantal_psychological_2012, yang_student_2024, joshi_writing_2025}. To complement this, we used the CSI to assess how engaging and meaningful students found the AI-supported writing process~\cite{cherry_quantifying_2014, ivcevic_artificial_2024, gero_sparks_2022}.
 By understanding engagement, we gain insight into whether students remain engaged in the writing process when ChatGPT performs cognitive work, a potential risk to learning. Together, these measures reveal how the use of GenAI affects students' sense of ownership and engagement in writing tasks.


 Our work contributes to understanding GenAI in writing education by providing four key contributions:
\begin{itemize}
    \item  Empirical, data-driven understanding of how college students use GenAI while writing, grounded in \citeauthor{flower_cognitive_1981}'s Cognitive Process Theory of Writing.
    \item  Insights into how user characteristics (e.g., age, race, gender, writing efficacy) relate to distinct GenAI usage patterns during writing. 
    \item  Understanding of how students' essays and their perceptions of their writing vary across usage patterns. 
    \item  A publicly available dataset of students’ GenAI interactions (queries and responses) paired with fine-grained editor interaction traces (keystrokes, copy–paste actions) from an argumentative essay writing task. 

\end{itemize}
 
The results revealed distinct patterns of ChatGPT usage, which we categorized each query into six groups, reflecting its role as an ideation partner, editor, ghostwriter, or some combination of these. We also identified a mode of vibe writing in which a subset of students compose essays by instructing ChatGPT to generate desirable outputs. Our findings indicate that SEWS predicts the frequency of ChatGPT use, and that students who rely on ChatGPT for planning or reviewing reported a comparable sense of ownership over their work, results that may raise concerns for educators. These insights---along with our publicly available dataset---provide educators and researchers with empirical resources to inform future GenAI policies, enabling them to examine how students' GenAI use may influence learning outcomes.


\section{Related Work} 
Research on the different aspects of intelligent assistants powered by LLM-based generative AI has been conducted in a variety of fields. 
In this section, we discuss GenAI's impact on education, AI writing support tools, and general attitudes surrounding GenAI tool usage.

\subsection{Mitigating the Impact of Generative AI in Education}

With the release of tools like ChatGPT, GenAI has become widely accessible as a conversational agent capable of responding to text, image, and audiovisual queries.  A 2024 Common Sense Media survey reported that 70\% of US teens---many soon entering university---had used generative AI, with more than half applying it to academic writing tasks such as idea generation and assignment completion~\cite{madden2024dawn}. In university admissions, students feel pressured to use AI because they think their peers use it ~\cite{fitzsimmons_pressure_2025}.

This rapid spread of GenAI is raising concerns among educators. The heavy reliance on GenAI can hinder critical thinking and self-evaluation, lower students’ confidence, and encourage uncritical acceptance of AI output, which risks amplifying inherent biases and reinforcing discriminatory viewpoints~\cite{jakesch2023co, holmes2023guidance}. 
Using LLMs may reduce mental effort~\cite{STADLER2024108386} and overall brain activity~\cite{kosmyna_your_nodate} and may result in student work that lacks depth~\cite{STADLER2024108386}. 
As students offload tasks to ChatGPT, they may hinder their own learning process by limiting the skills they develop. 

Additionally, there are ethical concerns focusing on plagiarism and cheating~\cite{tan_is_2024, kasneci_chatgpt_nodate, cotton_chatting_2023}. Studies show that AI-generated text often evades plagiarism detectors~\cite{orenstrakh_detecting_2024, quidwai_beyond_2023, zeng2024detecting}, although new detectors demonstrate improved accuracy~\cite{quidwai_beyond_2023}, and log file analysis has proven reliable for identifying unique writing patterns~\cite{schneider_detecting_2018}.
However, claiming GenAI misuse without definitive proof is risky and can lead to legal and ethical challenges~\cite{mndaily2024ai}.
Educators also worry about grading fairness, the difficulty in detecting ChatGPT use, and the broader risk of students losing opportunities to gain knowledge through active engagement~\cite{doi:10.1080/14703297.2023.2190148, AlAfnan2023ChatGPT}.
 
Researchers have examined the policies and regulations that govern LLM-powered tools in education. Some advocate strict policies and regulatory frameworks~\cite{adams_artificial_2022, cotton_chatting_2023, halaweh_chatgpt_2023, biswas_role_2023, sok_chatgpt_2023}, while others argue that adoption is inevitable and can be beneficial when guided appropriately~\cite{wang_exploring_2023, barrett_not_2023, bower_how_2024}. 
Educators face the challenge of integrating GenAI into their classrooms. Students often treat AI tools as collaborators in complex problem solving, with researchers documenting various use cases of AI in education~\cite{liu_teaching_2024, han_recipe4u_nodate, barrett_not_2023, bower_how_2024}.
Prior work highlights both benefits and challenges, offering strategies for effective classroom integration~\cite{baidoo-anu_education_2023, wang_effect_2025, song_enhancing_2023, park_promise_2024}. For example, \citeauthor{park_promise_2024} identified the strengths and weaknesses of ChatGPT with students and stakeholders, providing design ideas for classroom use~\cite{park_promise_2024}. Similarly, \citeauthor{harvey_dont_2025} recommends encouraging students to use ChatGPT for support when stuck, rather than as a replacement for problem-solving~\cite{harvey_dont_2025}, while \citeauthor{jeon_large_2023} propose strategies to foster complementary relationships between students, teachers, and AI~\cite{jeon_large_2023}. 
Research in HCI and education further emphasizes how LLMs can be integrated to deepen engagement and support meaningful learning experiences~\cite{hwang_review_2021, kasneci_chatgpt_nodate, shibani_ai-assisted_2024}. 

Although proper integration of GenAI may open a new avenue for education, our understanding of students' actual GenAI usage remains limited, with few studies focusing on native English speakers. Some researchers have investigated students' motivations for using tools like ChatGPT~\cite{black2025university, Skjuve_2024, ammari2025students, wasi_llms_2024}, but these studies often rely on self-reported data or query histories submitted by self-selected participants. As a result, they offer only a partial view of how students interact with GenAI. Even when query data is available~\cite{ammari2025students}, it often lacks the academic context in which the questions were asked, and instructors typically do not have access to it, making it difficult to comprehensively assess the educational impact of GenAI. Additionally, gender and ethnicity influence awareness and understanding of ChatGPT~\cite{cachero_gender_2025, grassini_gender}, potentially limiting which students turn to AI for support.

 There is existing research providing comprehensive datasets of LLM usage~\cite{han_recipe4u_nodate, lee_coauthor_2022, liu_detectability_2024, laskar_systematic_2023, chatterji_how_2025, wang_exploring_2023}. 
These datasets cover a variety of contexts, ranging from academic programming and problem-solving tasks~\cite{wang_exploring_2023, laskar_systematic_2023} 
to professional work environments~\cite{chatterji_how_2025}.
Most of these datasets only provide snapshots of the process, for example, only collecting queries.
 RECIPE4U is one of the rare exceptions—developed by \citeauthor{han_recipe4u_nodate}---that collected both editor states and queries during student---ChatGPT interactions as students completed a guided essay-revision task in an English as a Foreign Language (EFL) class~\cite{han_recipe4u_nodate}. Because this dataset was collected in an EFL context, it primarily focuses on language-related queries and revision behaviors.
We extend this line of work by moving beyond the EFL context and providing a detailed, process-level trace of students’ interactions and writing behaviors, including keystroke and copy–paste events. This low-level data enables the reconstruction of the complete writing process and supports deeper analyses of the relationships between AI usage patterns and writing outcomes.
This dataset will support a deeper understanding of the discourse around the risks of and responses to GenAI in education, while underscoring the need to examine its role in specific domains such as writing education~\cite{han_recipe4u_nodate, knight_acawriter_2020, shibani_ai-assisted_2024, kasneci_chatgpt_nodate, liu_teaching_2024}.

\subsection{AI Writing Assistants in Education}

Communicating ideas through writing is a critical skill across disciplines as it fosters critical thinking, analysis, and synthesis~\cite{condon2004assessing, wade1995using, emig1977writing}. To support this learning process, AI has long been incorporated into writing education through automated feedback systems and tutoring platforms~\cite{knight_acawriter_2020, shibani2019contextualizable, 10.1145/3411764.3445683, escalante2023ai, crompton_artificial_2023}.

Prior work demonstrates that AI writing assistants can improve the writing process and support learning. Commercial tools like Grammarly help users avoid plagiarism and improve writing quality, especially for EFL learners~\cite{grammarly, dong_using_2021, 
koltovskaia_student_2020}, though students often underuse capabilities~\cite{huang_effectiveness_2020}. Educational systems like AcaWriter demonstrate that automated rhetorical feedback can support academic writing and improve essay quality~\cite{knight_acawriter_2020}.
RECIPE extends this by integrating ChatGPT into EFL classrooms, showing that interactive revision support can improve performance and satisfaction compared to traditional instruction~\cite{han_recipe4u_nodate}.  Similarly, CoachGPT provides scaffolding-based support for essay planning and reviewing, with positive student perceptions~\cite{chen_coachgpt_2025}. LegalWriter demonstrates benefits in interdisciplinary legal writing contexts, showing that LLM-based feedback improves writing quality and student learning outcomes~\cite{weber_legalwriter_2024}. In parallel, Langsmith explored how Japanese learners use AI translation to complete writing tasks, finding that students relied heavily on the tool and focused more on text quality than manual translation~\cite{ito-etal-2020-langsmith}.

 To understand how these tools integrate into writing, researchers have turned to established writing frameworks, particularly \citeauthor{flower_cognitive_1981}' cognitive process theory.
  This writing process, a widely used writing framework, involves dynamically and recursively switching between three basic writing processes: \textit{Planning}, \textit{Translating}, and \textit{Reviewing}~\cite{flower_cognitive_1981}.
  This model is grounded in the key point that ``\textit{writing is best understood as a set of distinctive thinking processes}.'' The theory emphasizes think-aloud protocols, which capture a detailed record of a writer’s cognitive processes during composition, rather than relying on post hoc introspection reflecting what writers believe should have happened. Drawing on this writing model and combining systematic literature review with user studies, these frameworks reveal patterns in how writers integrate AI across different phases and contexts~\cite{reza_co-writing_2025, lee_design_2024, shibani_ai-assisted_2024, goldi_support_2024}. This framework provides a way for how students integrate AI across writing phases, which we adopt in our analysis.

  Beyond frameworks, recent research has examined how individual characteristics shape AI engagement in writing contexts.
 \citeauthor{joshi_writing_2025} investigates how well prompting strategies play into perceived ownership of essays~\cite{joshi_writing_2025}, finding that ownership is impacted by the length and detail provided to ChatGPT inquiries.  Other work suggests that perceived ownership is dependent on the specific writing tasks~\cite{wasi_llms_2024}.
 Using thematic coding grounded in cognitive writing theory, researchers have identified three key elements that shape AI engagement: a participant's personal values, their relationship with AI, and the different integration strategies, revealing how individual differences influence usage patterns~\cite{guo_pen_2025}.

 HCI research has examined professional and creative writers' use of AI, revealing that they typically want assistance in translating ideas and reviewing their essays~\cite{10.1145/3635636.3656201, gero_social_2023}. These writers show varied expectations but general appreciation for ChatGPT's ability to generate unexpected, sometimes inspiring output~\cite{gero_social_2023, wan_it_2024, lee_coauthor_2022}. However, less work has focused on students in authentic educational contexts ---a critical gap as students are learning proper writing strategies. Research with students shows they engage more when using ChatGPT for reviewing processes~\cite{goldi_support_2024}, and while LLM usage does not necessarily speed up writing, students report increased time spent engaging with their work and perceive higher quality~\cite{goldi_support_2024, wilbers_overall_2024}.

Despite these promising findings, practitioners have raised concerns about the negative impacts of GenAI that are difficult to regulate, including weakened critical engagement~\cite{herman2022end, marche2022college, Livingstone2024, EKE2023100060, scarfe2024real, mndaily2024ai}. 
 Furthermore, research also shows limits to AI writing quality~\cite{chakrabarty_art_2024, romoff_role_2025}; AI-generated essays pass evaluation criteria 3-10 times less frequently than professional human-written essays~\cite{chakrabarty_art_2024}. These concerns highlight the importance of understanding how students use AI across different writing phases.

While this growing body of work provides valuable insights into AI adoption, perceptions, and writing processes, most observational studies of AI-assisted writing focus on professional or creative writers in controlled settings. Less attention has been paid to how students engage with GenAI during authentic academic writing tasks, and how their individual characteristics---such as writing ability, AI literacy, or personal values---shape these interactions.  This gap motivates our study, which directly observes students using ChatGPT for essays through the lens of \citeauthor{flower_cognitive_1981}' framework to understand typical usage patterns, individual variation, and how students integrate AI across writing phases.

\subsection{Attitude and Usage of AI}

    
  As artificial intelligence is a rapidly expanding field, there is constant growth in understanding its adoption and perceived quality. 
  In a 2023 study, \citeauthor{chan_ai_2023} found that students were more likely than teachers to adopt new AI tools and had a more open-minded attitude about their use
  Other research looks at AI literacy and interest, talking about the importance of fostering positive attitudes towards AI~\cite{bewersdorff_ai_2025}. 
  \citeauthor{ayanwale_teachers_2022} studies this through teachers' intention to teach using AI~\cite{ayanwale_teachers_2022}, while others performed thematic analysis or interviews to get an understanding of perception~\cite{shoufan_exploring_2023, ali_impact_2023}. 
Other researchers study perceptions of AI quality and trustworthiness. Interestingly, users tend to accept AI suggestions regardless of agreement~\cite{bhat_interacting_2023}, while others examine what makes responses perceived as helpful or unsuitable~\cite{ma_who_2023, bucinca_trust_2021}. They also discuss ways in which the user can improve the prompts to increase the likelihood of getting a good response. \citeauthor{zamfirescu-pereira_why_2023} expanded this idea, developing an accessible prompt engineering tool that doesn't require background knowledge of AI to use~\cite{zamfirescu-pereira_why_2023}. 
   Other research discusses prompting strategies and how gender plays a role in AI acceptance~\cite{sawalha_analyzing_2024, grassini_gender, cachero_gender_2025}, finding that people treat GenAI like humans, get better responses using revised prompts, and that women have more hesitation towards AI.

While this growing body of work provides valuable insight into AI adoption, literacy, and perceptions, it focuses largely on attitudes rather than the ways people, especially students, actually use GenAI in practice. In particular, little is known about how students engage with GenAI during authentic learning activities or how their individual characteristics shape these interactions. Understanding student usage requires moving beyond general attitudes and quality perceptions to observe actual interactions in academic contexts. This gap motivates our study, which directly observes how students use ChatGPT in academic writing to understand their interaction patterns and implications for learning.

\section{Method}
To understand the usage patterns students have with ChatGPT, we conducted an online study designed to capture these patterns and examine their relationship with other factors: the students’ background, the resulting essay, and their perceptions towards AI. We introduce the system developed for data collection and outline the methodological approach used for our qualitative and quantitative analyses.

\begin{figure*}[ht!]
    \centering
    \includegraphics[width=1.1\linewidth]{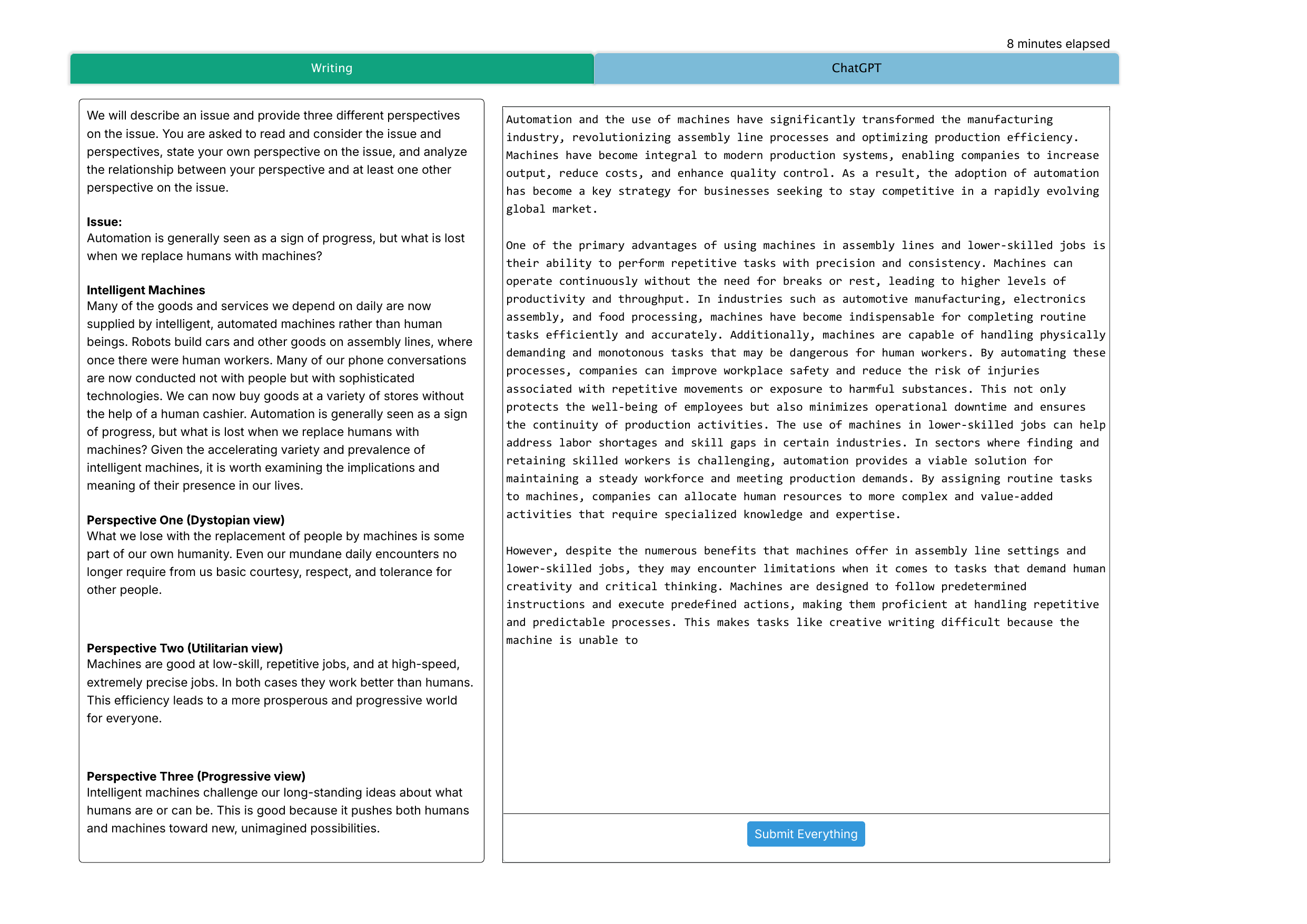}
    \hspace{-1.5cm}
    \vspace{-1.0cm}
    \caption{The editor view of the website}
    \Description{Split-pane writing interface with ACT prompt on left and text editor on right. Dual-tab navigation enables switching between writing and ChatGPT modes. The text editor contains writing in progress for an essay.}
    \label{fig:writ_page}
\end{figure*}

\subsection{Instrument Development: Writing Platform + ChatGPT Development}
To understand how students use ChatGPT, we developed a platform that tracked their queries and the corresponding responses. Since ChatGPT is an independent app, we built a system that integrates an \textit{in-house ChatGPT} — referred to simply as ChatGPT from this point — within the writing platform, using the default OpenAI API (model 3.5-turbo) to record user interactions. This tool enabled us to collect three types of data: students' queries to ChatGPT, ChatGPT's responses, and keystroke-level recordings of their writing process, which allowed us to analyze how students incorporated ChatGPT responses into their essays.
Our application has two main features: a plain text editor for essay writing and access to ChatGPT. The web application emulates ChatGPT's functionality to replicate its experience as closely as possible. 

The first tab (\autoref{fig:writ_page}) of our application is a writing platform where participants were asked to respond to an essay prompt in the text editor. The editor recorded all input operations and their sequence, including insertions, deletions, text selection, copy, cut, and paste events. We also recorded the timestamps of each operation to determine when each edit was made. With this, we were able to observe and analyze participants' writing processes, using timestamps to track how they alternated between the editor and the in-house ChatGPT and how they integrated ChatGPT responses into their writing (e.g., pasted text). This data was sent to a server as it was generated, and these features were implemented using the CodeMirror 5 API and the CodeMirror-Record files~\cite{Jisuanke2023}. 

\begin{figure*}[h!]
    \centering
    \includegraphics[width=\linewidth]{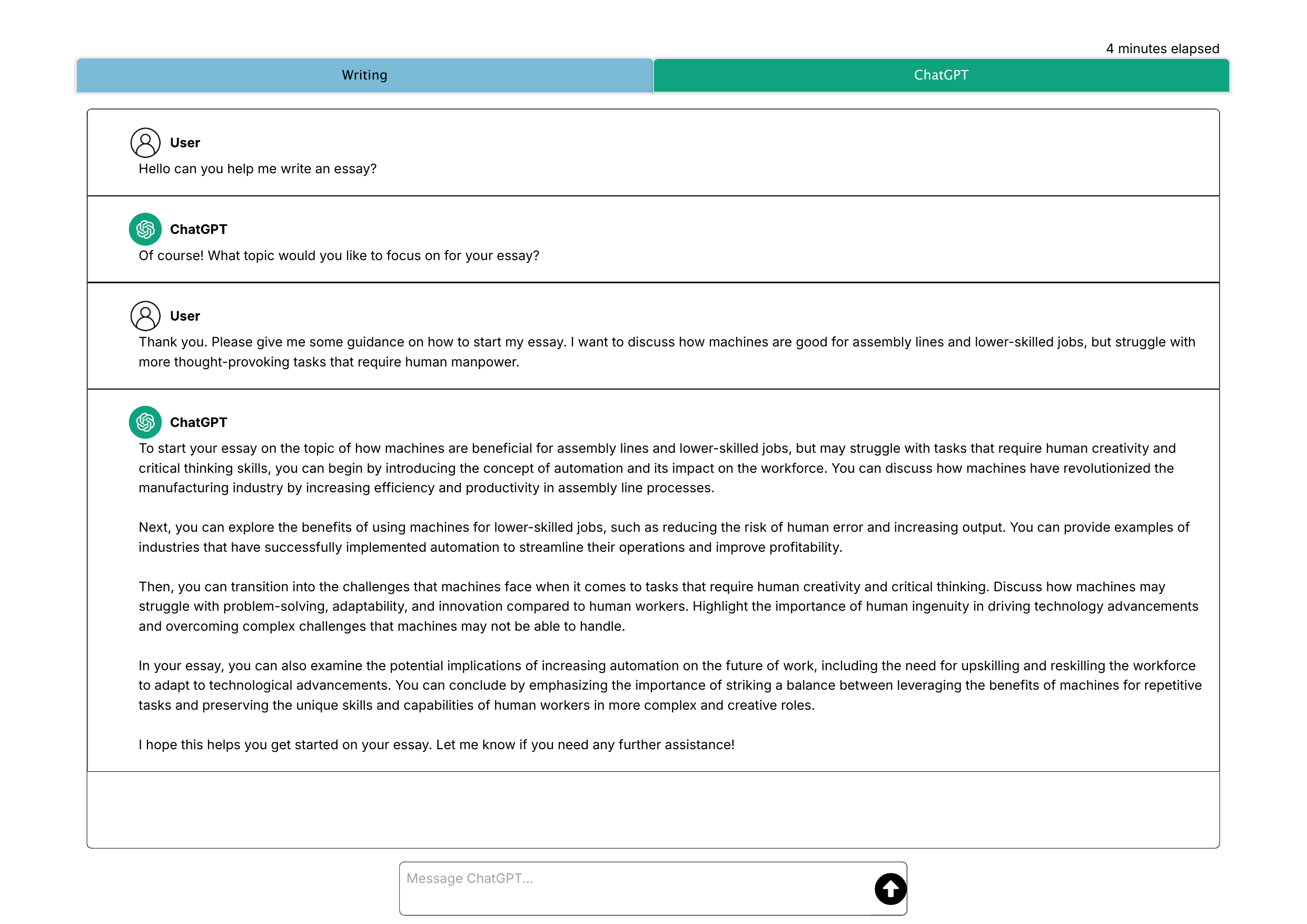}
    \caption{The screencapture of the in-house ChatGPT provided to participants}
    \Description{ChatGPT interface with conversation history display above query input field. Tab navigation enables switching between ChatGPT and writing modes.}
    \label{fig:gpt_page}
\end{figure*}

To track how users interact with ChatGPT, we implemented a custom version of ChatGPT using the OpenAI API, as shown in \autoref{fig:gpt_page}. We chose to simulate browser tabs to give participants the impression that ChatGPT was available to them in a separate window, requiring them to switch tabs if they wanted to use it. This design choice mirrors practical usage, as opposed to displaying ChatGPT side by side with the writing platform, which could artificially encourage their use. Participants were allowed to ask any questions to ChatGPT, and we did not pre-prompt the system (e.g., assigning it the role of a writing assistant) so that its behavior would closely resemble the standard ChatGPT experience.  We also allow for copy/paste events between each window, allowing participants to bring information between both tabs. We recorded all queries and their timestamps to analyze how and when ChatGPT was prompted for assistance during the writing process. 


\subsection{Study Procedure}
Before writing an essay, we asked participants to complete a pre-study survey created in QuestionPro to collect basic demographic information. The survey also included two standard questionnaires: the Technology Acceptance Model (TAM)~\cite{davis_perceived_1989}, which we adapted for ChatGPT, and a Self-Efficacy for Writing (SEWS) questionnaire~\cite{bruning2013examining}. This questionaire is shown in Appendix \ref{app:pre_survey}. 

The Technology Acceptance Model (TAM) is a well-established framework used to understand how users come to accept and use technology. TAM has two subscales: (1) perceived usefulness of technology (TAM PU), which refers to the degree to which a person believes that using a particular technology will enhance their job performance or improve productivity, and (2) perceived ease of use (TAM PEOU), which reflects how easy a technology is to use, based on the idea that users are more likely to accept a technology if they find it easy to operate. The Self-Efficacy for Writing Scale (SEWS) measures students’ confidence in their writing abilities across different tasks and contexts. This construct is grounded in influential motivation and writing theories and also accounts for behaviors such as help-seeking. Students with lower self-efficacy may be more inclined to use tools such as ChatGPT to reduce the cognitive load~\cite{woodrow_college_2011, bruning2013examining}. Together, these two constructs provide complementary insights: SEWS reflects internal beliefs about writing competency, while TAM captures perceptions of the tool itself. By incorporating both, we examine which factors can serve as predictors of GenAI usage, accounting for how students may or may not use ChatGPT in their writing.

The participants were then redirected to our writing-ChatGPT platform to begin the essay task. We used a sample writing prompt from the American College Testing (ACT), as most college students applying to US universities are familiar with this type of assignment. The prompt addressed the issue of automation replacing humans with machines and included three perspectives on the topic. Participants were asked to present their own perspective and analyze how it relates to at least one of the perspectives provided (shown in Figure \ref{fig:writ_page}). The complete prompt is included in Appendix~\ref{app:writing}.
We asked participants to spend approximately 30 minutes on the essay, as the ACT exam allows a maximum of 40 minutes for the essay response. During the study, they were neither encouraged nor discouraged from using ChatGPT. The study was advertised as ``a study investigating essay writing and ChatGPT." On the interface, the participants were instructed: ``If you wish to use ChatGPT, please click the ChatGPT tab and ask questions. Do not use ChatGPT in your browser; use the one we provided." They were further instructed to write the essay as if it were ``a class assignment that would be submitted for a grade."
 We did not impose any policy restricting ChatGPT usage, allowing participants full autonomy in deciding whether and how to use the tool.

    
After submitting their essays, participants completed two additional questionnaires to reflect on their writing experience. First, we sought to determine whether students felt the essay was truly ``theirs" and whether reliance on ChatGPT influenced that perception. To measure perceived ownership (PO) of the written artifact, we used a validated questionnaire~\cite{avey_psychological_2009, chantal_psychological_2012, vandewalle_psychological_1995}. Second, we used the Creativity Support Index (CSI) to assess how well ChatGPT supported their creativity~\cite{cherry_quantifying_2014}. 
From CSI, we took the Collaboration subscale questions out as this scenario does not involve any collaboration with others. 
These measures provide insight into students’ perceptions of ChatGPT in the context of academic writing.


\subsection{Recruitment}
For recruitment, we posted our survey on various university mailing lists, targeting both undergraduate and graduate students. In addition, we recruited participants through Prolific, an online crowd-sourcing platform, with the screening criterion of being a college student in the United States. All participants were entered into a raffle for the chance to win a \$10 gift card, with a winning odds of 1 in 5. In total, we recruited 77 participants. For participants’ ages, we used the following age bands: 18–24 (n=62), 25–34 (n=10), 35–44 (n=2), 45–54 (n=1), 55–64 (n=2). Of the 77 participants, 34 identified as women, 1 as nonbinary, and 42 as men. The racial distribution was as follows: 37 White/Caucasian, 24 Asian/Pacific Islander, 7 Black or African American, 5 Hispanic, and 4 Other.

\subsection{Qualitative Analysis of ChatGPT Queries}
\label{sec:query-cats}
To analyze the queries sent to ChatGPT, we coded all participant queries 
 into categories inspired by \citeauthor{flower_cognitive_1981}’ Cognitive Process Theory of Writing~\cite{flower_cognitive_1981}. 
 In this study, we treat the queries that students submit to ChatGPT as an alternative source of think-aloud data, as these queries manifest their ongoing cognitive processes. While ChatGPT cannot gather all of the data a true think-aloud study can, this approach captures authentic student usage and explores how cognitive processes across writing phases vary with individual characteristics. In addition, having independent coders categorize each query based on its semantic content mirrors the real-world practice in which instructors must assess the nature and severity of GenAI use from the questions students ask, rather than relying on students to articulate their underlying intentions.


\begin{figure*}[ht!]
    \centering
    \includegraphics[width=\linewidth]{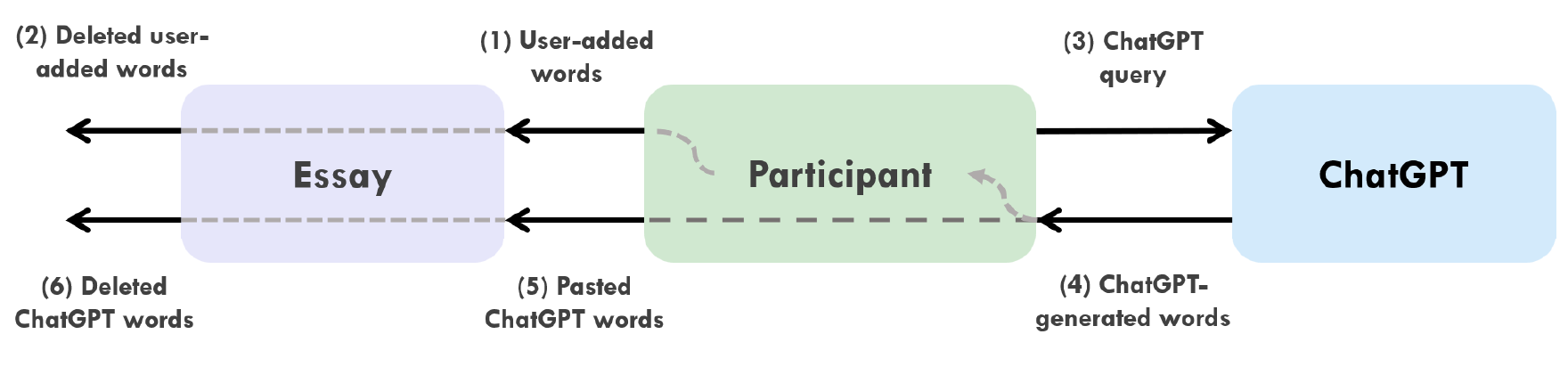}
    \caption{Data Flow Diagram between Editor, Participant, and ChatGPT}
    \Description{Data flow diagram showing actions made in by participants. }
    \label{fig:dataflow}
\end{figure*}

In the following subsections, we provide a detailed description of each category and outline the subcategories that have been identified within it. Coding and categorization were performed independently by two authors, followed by multiple discussion sessions to reach agreement  (Cohens $\kappa = .89$; see \autoref{sec:irr}).

\subsubsection{\textbf{Planning (P)}}
According to \citeauthor{flower_cognitive_1981}'s model, the goal of \textit{Planning} is defined as ``\textit{to take information from the task environment and from long-term memory and to use it to set goals and to establish a writing plan to guide the production of a text that will meet those goals}''~\cite{flower_cognitive_1981}. Main activities within planning include collecting information, generating and organizing ideas, setting goals or outlining the writing structure~\cite{flower_cognitive_1981}. This category, therefore, included queries such as asking for examples, seeking information, or requesting help in structuring the essay. Any query explicitly asking ChatGPT to generate essay text or write portions of the essay were not identified as planning and classified into a different category.

\subsubsection{\textbf{Translating (T)}}
The second category derived from the model is \textit{Translating}, defined as the process of turning ideas into text~\cite{flower_cognitive_1981}. Queries were classified as translating when they included both a request to generate text for the essay and sufficient context about the desired content. Requests to generate portions of the essay larger than a paragraph were excluded from this category and instead classified into the \textit{All} category, as this includes planning and reviewing activities.
    
\subsubsection{\textbf{Reviewing (R)}}
The \textit{Reviewing} category included any query that asked for evaluations or revisions of existing text, aligning with the two subprocesses of reviewing in Flower and Hayes' model~\cite{flower_cognitive_1981}. Queries involving evaluation ranged from seeking targeted feedback on written text to requesting a score or grade. Queries involving revision included requests to fix simple spelling and grammar errors, as well as more complex tasks, such as rewriting sections of an essay in a particular style. In other words, any query in which participants supplied original text and requested an evaluation or rewrite---without significantly altering the overall theme or viewpoint---was classified as \textit{Reviewing}.

\subsubsection{\textbf{All (A)}}
The \textit{All} category corresponded to using ChatGPT to make a request that involves \textit{all} three activities, Planning, Translating, and Reviewing, to ChatGPT.
Such queries involve generating the entire essay or a portion of it (e.g., a paragraph).
Finally, not all queries fit the four main categories. Some were unrelated to the writing task, as there were no constraints on the types of queries that participants could submit. For example, we identified a code for \textit{providing feedback}, where a participant evaluated ChatGPT’s response and offered feedback (e.g., typing ``That's a great idea.''), which is not tied to any specific process of writing. We do not discuss these types of messages in this paper, as our focus is on writing-related queries.

\subsubsection{\textbf{Inter-rater Reliability}}
\label{sec:irr}

To validate our qualitative coding results, two researchers independently categorized each ChatGPT query using our coding framework (P, T, R, A). Our analysis involved categorical classification into four distinct codes, and the resulting Cohen Kappa score ($\kappa = .89$) indicates strong inter-rater agreement~\cite{mchugh2012interrater}, demonstrating both the clarity and consistency of our codebook definitions and supporting the validity of the subsequent quantitative analysis based on the four categories.

\subsubsection{\textbf{Finer-grained Coding}}
 
In addition to the four categories above, we coded each query using a finer-grained codebook that we developed in an inductive manner. 
These codes were defined \textit{within} one of the four process categories, so for example, the code \textit{Proofread} (RE01 in Table~\ref{tab:reviewing}) falls under the Reviewing category.
Two authors of the paper started out developing the codebook independently, then met to compare their coding schemes and revise them through discussion until they reached agreement. 
The finer-grained codes for each category are presented in the Results section (Tables~\ref{tab:planning},~\ref{tab:translating}, ~\ref{tab:reviewing}, and ~\ref{tab:all}).
The authors assigned a code based on the semantic meaning of the query itself, rather than the latent intention that can be inferred from subsequent actions (e.g., how the generated text was used afterwards), thereby minimizing the subjectivity of coders in the coding process.

\subsection{Quantitative Analysis}

\subsubsection{Essay Writing Trace}
The recording features tracked each user's input and stored it in our database with timestamps. In general, it provided comprehensive keystroke-level data capable of reproducing each writer’s writing process and interactions with ChatGPT. \autoref{fig:dataflow} illustrates the overall data flow in terms of word count. For example, we examined how ChatGPT responses contributed to the writing process by comparing the generated responses with the text participants pasted into the editor afterward, represented as ``Pasted ChatGPT words'' in \autoref{fig:dataflow}. The following list provides examples of metrics calculated for each participant:

\begin{itemize} [left=.3cm, labelsep=0.15cm]
    \item Number of queries made (per category: P, T, R, A) for the essay (\autoref{fig:dataflow}-$(3)$)
    \item Number of words manually entered (\autoref{fig:dataflow}-$(1)$)
    \item Number of words copy-pasted from ChatGPT into the essay (per query category: P, T, R, A) (\autoref{fig:dataflow}-$(5)$)
    \item Number and word count of Copy/Cut/Paste events in the Editor, Prompt, or ChatGPT query textbox
    \item Final number of ChatGPT-generated words in the essay (\autoref{fig:dataflow}-[$(5)-(6)$])
    \item Final number of participant-written words in the essay (\autoref{fig:dataflow}-[$(1)-(2)$])
\end{itemize}

\noindent We used these metrics to gain insight into how users interact with GenAI and how their use relates to other constructs~\footnote{Note that these observed metrics have limitations and cannot fully capture users' cognitive and behavioral processes. For example, if a participant writes words manually, we cannot determine whether those words were generated from their own memory and knowledge or derived from a ChatGPT response and rephrased in their own way. In such cases, the words are categorized as user-added, since they were typed rather than copy-pasted (depicted as a gray dotted line passing through the participant in \autoref{fig:dataflow}-(1)).}.

\begin{figure*}[h!t]
    \centering
    \begin{subfigure}[b]{0.49\textwidth}
        \centering
        \includegraphics[width=\textwidth]{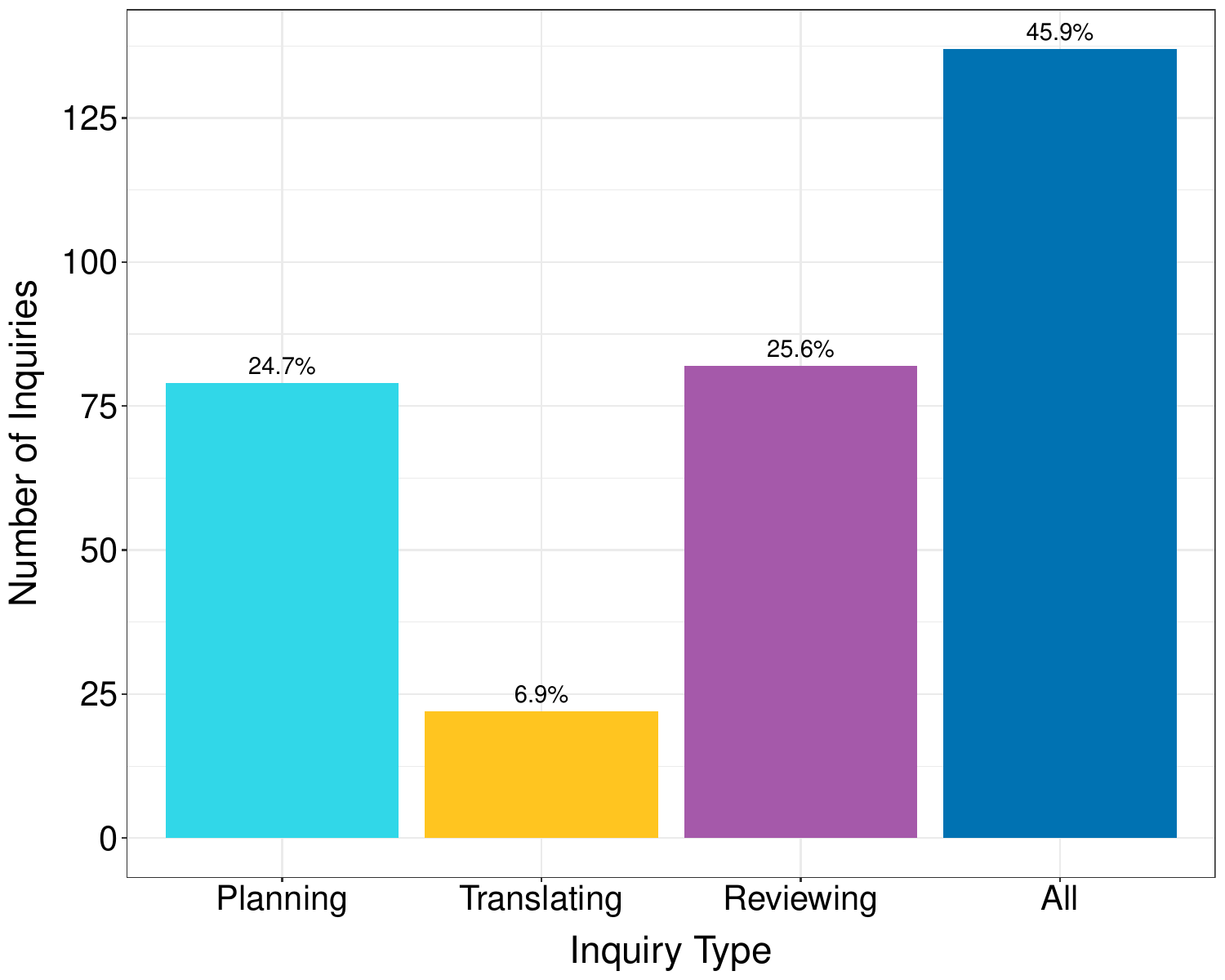} 
        \caption{Number of queries asked per group}
        \Description{Bar graph showing query distribution across four categories ranging zero to 125. These categories are determined by our qualitative coding. All category contains the highest percent of total queries at 45.9 percent.}
        \label{fig:countquery} 
    \end{subfigure}
     \hfill 
    \begin{subfigure}[b]{0.49\textwidth} 
        \centering
        \includegraphics[width=\textwidth]{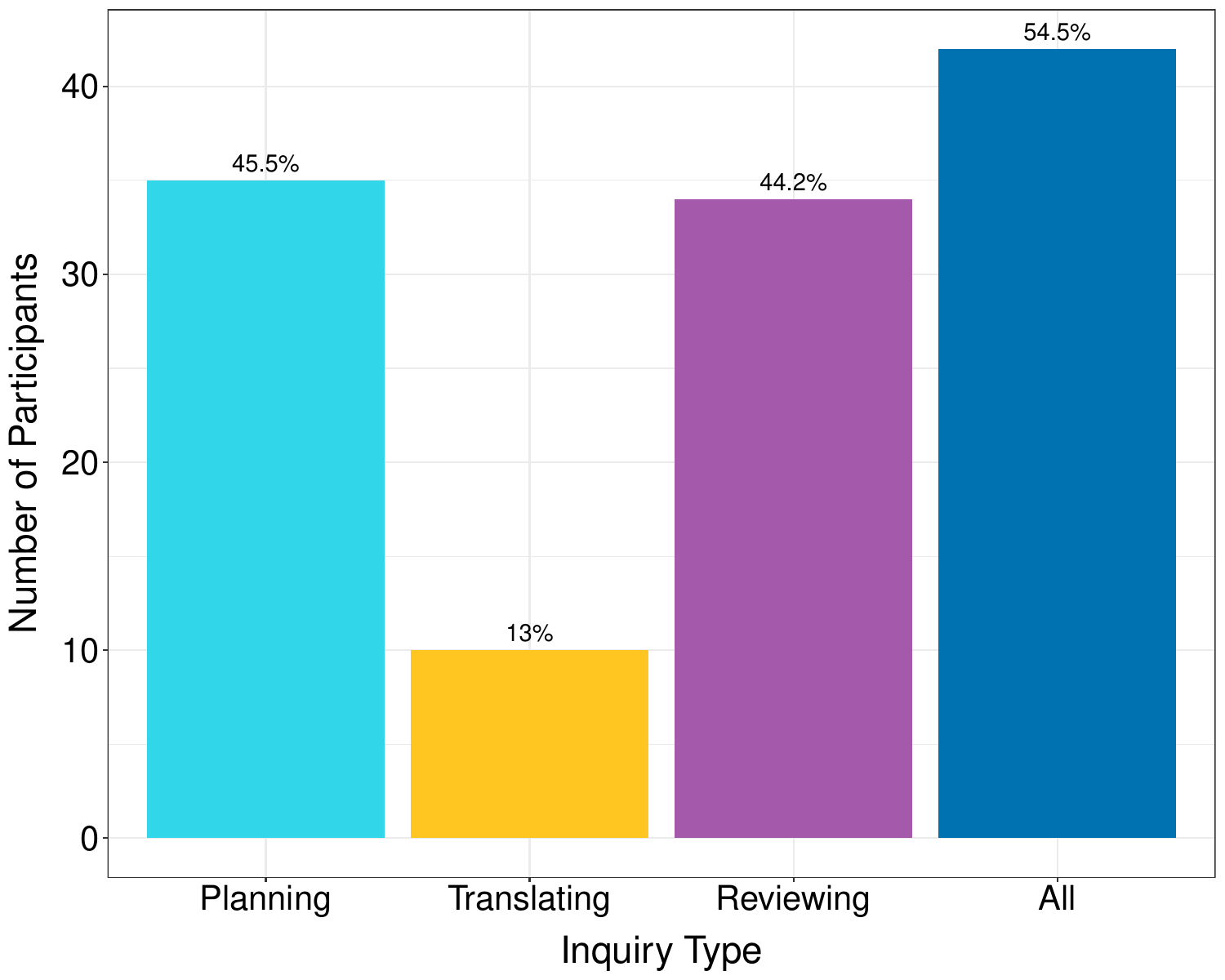}
        \caption{Number of participants that asked each query}
        \Description{Bar graph showing participant engagement across four categories ranging zero to 40. These categories are determined by our qualitative coding. 54.5 percent of participants engaged with All category, our highest engagement.}
        \label{fig:partinquiry} 
    \end{subfigure}
    \caption{The query count for each category and the number of participants who used such a query}
    \Description{Two bar graphs showing the distribution of the queries across the categories detailed in our Qualitative Coding. The leftmost graph shows inquiry type distribution across queries. The rightmost graph shows inquiry type across participant count.}
    \label{fig:query_info} 
\end{figure*}

\subsubsection{Data Analysis for RQ2}


For RQ2, we examined whether an individual's background, including Self-Efficacy for Writing (SEWS) and Technology Acceptance Model (TAM) scores, could predict ChatGPT usage. We ran a generalized linear model (GLM) using the \texttt{glm} function in R. A Poisson model was chosen because the predicted values were counts (e.g., the number of ChatGPT queries or the number of words written by a participant). For example, the relationship between the predictors and the expected count of ChatGPT queries ($\mu_i$) for participant ($i$) can be expressed as follows:

\[
\begin{aligned}
\log(\mu_i) &= \beta_0 + \beta_1 \cdot \text{SEWS}_i + \beta_2 \cdot \text{TAM PU}_i + \beta_3 \cdot \text{TAM PEOU}_i \\
&\quad + \beta_4 \cdot \text{Gender}_i + \beta_5 \cdot \text{Race}_i + \beta_6 \cdot \text{Age}_i 
\end{aligned}
\]

\noindent
where:
\begin{itemize}[left=.3cm, labelsep=0.15cm]
    \item $\text{Gender}_i = 1$ for Men, $0$ otherwise
    \item $\text{Race}_i = 1$ for White, $0$ otherwise
    \item $\text{Age}_i = 0$ for ages $18\text{--}24$, $1$ for $25\text{--}34$, $2$ for $35\text{--}44$, $3$ for $45\text{--}54$,  $4$ for $55$ or older
\end{itemize}
\noindent The model uses a log-link function to relate the expected query count for each type (P, T, R, A, and their total) per essay to the predictor variables.

\subsubsection{Readability scores for RQ3}
 Additionally, we looked at the complexity of the essays using readability scoring. We chose to use both the Flesch-Kincaid Grade Level and Dale-Chall Readability Score. The Flesch-Kincaid Grade Level is one of the more commonly used metrics, estimating the U.S. grade level required to comprehend the text, with higher scores indicating greater complexity~\cite{flesch_new_1948, klare_measurement_2000}. The Flesch-Kincaid metric uses average sentence length and syllable count per word to determine text complexity~\cite{flesch_new_1948}. 
We also used the Dale-Chall Score as it has been found to be more accurate than other readability metrics, including Flesch-Kincaid~\cite{klare_measurement_2000}. 
The Dale-Chall Score evaluates readability based on sentence length and word difficulty~\cite{dale_formula_1948}, complementing the Flesch-Kincaid Grade Level by using a vocabulary-based assessment to capture semantic complexity.

\section{Results}

\subsection{RQ1: Understanding Students' ChatGPT Queries for Essay Writing}
%
 
In understanding how students' ChatGPT usage fit within their writing processes, we analyzed the queries they sent to ChatGPT.
As discussed in \ref{sec:query-cats}, we grouped queries into four main categories---Planning, Translating, Reviewing, and All---developing finer-grained codes for a more detailed analysis.

In total, participants sent $361$ messages to ChatGPT, and we identified $26$ unique codes. Of these, $320$ messages fell into one of the four main categories. \autoref{fig:query_info} shows the number of query messages that were categorized into each type and the number of participants who had such a query at least once.

Messages that were not categorized (41) were typically incomplete and excluded to avoid redundant counting. For example, a participant might ask ChatGPT to review their essay without pasting the text, followed by a repaired query that included it; in such cases, only the repaired query was categorized. A small subset of miscellaneous codes (Greeting, Appreciation, Jokes, and Clarifying ChatGPT’s capability) were identified but are not discussed in this paper.  With this coding scheme, we identify how the different processes of writing appear in the queries to ChatGPT, presenting frequency and subcategories describing how the process appeared in participant queries. Below, we present the codes within each category.

\subsubsection{\textbf{Planning}}

Planning queries refer to the process of ideating and deciding what to write, often occurring in the early stages of writing. Nearly half of the participants (35/77) used at least one Planning query while writing their essays. Among these participants, the average frequency was more than two queries each, and we observed diverse uses of ChatGPT related to Planning.

\begin{table}[h!]
    \centering
    \caption{\textbf{Planning} Query Types}
    \label{tab:planning}
    \begin{tabular}{c p{0.42\linewidth} c c}
            \toprule
Code & What did they ask ChatGPT to do? & Count & Participants (\%) \\
        \hline
        PL01 & Provide an answer to a question on a topic & 20 & 12 (15.6\%) \\
        PL02 & Provide examples & 16 & 12 (15.6\%) \\
        PL03 & Search for factual information & 18 & 8 (10.4\%) \\
        PL04 & Suggest an essay structure & 10 & 7 (9.1\%) \\
        PL05 & Expand on an existing idea & 7 & 7 (9.1\%) \\
        PL06 & Recommend topics to write about & 5 & 4 (5.2\%) \\
        PL07 & Help interpret the writing prompt & 4 & 4 (5.2\%) \\
        PL08 & Compare the essay to an alternative viewpoint & 3 & 2 (2.6\%) \\
        \bottomrule
    \end{tabular}
    \Description{Top Planning query codes ranked by frequency. Our most common code, PL01 - "Provide an answer to a question on a topic",  accounts for 20 instances across 12 (15.6 percent) of participants.}
\end{table}

The most common type of Planning query involved using ChatGPT as a tool for topic research, particularly for requesting examples (PL02) or retrieving factual information such as statistics (PL03). Examples of this type of query are provided below.
\begin{itemize}
    \item P49: \textit{Can you provide examples for a machine doing better at a task than a human would?} (PL02)
    \item P75: \textit{average unemployment rate in the 2000s} (PL02)
\end{itemize}

\noindent In this case, the students used ChatGPT much like an online search engine. At times, answers (e.g. statistics) were not provided because the model powering the in-house ChatGPT did not have web search functionality, and it refused to answer based on hallucination. However, when an answer was generated, it was often more convenient than using a traditional search engine, as participants did not need to navigate multiple result pages or collect information before selecting what to use in their essays.

Another common type of query involved simply asking a question related to a topic or taking writing suggestions (PL01, PL05, PL06 in Table~\ref{tab:planning}). In these cases, the queries resembled asking someone for their opinions rather than searching for information. Examples in this category are as follows:
\begin{itemize}
    \item P20: \textit{Give me reasons why automation is actually good for society to be able to progress} (PL01)
    \item P36: \textit{Automation is generally seen as a sign of progress, but what is lost when we replace humans with machines? [copied from the prompt]} (PL01) 
    \item P54: \textit{read this information about an issue and give some ideas for an essay I am going to write about it. [pasted writing prompt]} (PL06) 
\end{itemize}

\noindent Although this type of question does not explicitly ask ChatGPT to write the essay, it replaces the process through which a student might otherwise critically engage with the topic via contemplation or research. While P20 at least appeared to select one of the three perspectives provided in the prompt, P36 and P54 did not generate any questions or ideas of their own. Specifically, P36's query was a direct copy-and-paste of the writing prompt, and P54 explicitly asked ChatGPT to suggest ideas for what to write about.

Finally, rather than asking about the essay topic itself, some participants asked questions about how to \textit{structure the essay} (PL04). 
Even when students have an idea of what to write about, arranging and organizing those ideas into a coherent narrative is a critical aspect of writing, as it shapes how effectively the overall topic is conveyed to readers. Relying on ChatGPT for this step can represent a missed opportunity to develop one's own writing expertise~\cite{bereiter2013psychology}.

Overall, using Planning queries with ChatGPT replaced essential aspects of writing, such as forming an opinion on a topic, conducting research, and determining the overall direction of the essay (e.g., logic, flow, structure).        
    
\subsubsection{\textbf{Translating}}
Translating is a crucial process of the writing process that shapes the overall quality and rhetoric of an essay beyond its core ideas, influencing aspects such as tone, coherence, and persuasiveness~\cite{flower_cognitive_1981}. We labeled a participant query as Translating when the participant provided an idea and/or surrounding text from which the idea could be inferred and asked ChatGPT to generate text that they were going to use for writing; these queries primarily reflect generating text based on the given idea (shown in Table \ref{tab:translating}). 
Only 10 participants used ChatGPT to support the translation process of writing, making it the least common query type in our dataset.

\begin{table}[h!t]
\centering
\caption{\textbf{Translating} Query Types}
\label{tab:translating}
\begin{tabular}{c p{0.42\linewidth} c c}
\toprule
    Code & What did they ask ChatGPT to do? & Count & Participants (\%) \\
    \hline
    TR01 & Write a paragraph given an idea & 9 & 7 (9.1\%) \\
    TR02 & Complete incomplete paragraphs/sentences & 10 & 6 (7.8\%) \\
    TR03 & Write a sentence given an idea & 3 & 3 (3.9\%) \\
    TR04 & Suggest expression/word choice & 5 & 2 (2.6\%) \\
    \bottomrule
\end{tabular}
\Description{Top Translating query codes ranked by frequency. Our most common code, TR01 - "Write a paragraph given an idea",  accounts for 9 instances across 7 (9.1 percent) of participants.}
\end{table}  



The most common type of Translating query occurred when participants asked ChatGPT to write a paragraph (TR01) or a sentence (TR03) based on an idea they provided, a pattern observed with seven participants.
\begin{itemize}
    \item P04: \textit{Write a final intro sentence that explains what this paper is trying to do: aka ``this paper claims that ai is good because it exposes our inefficiencies'' but in a few sentences.} (TR01)
    \item P58: \textit{[A paragraph pasted] make this paragraph stronger.} (TR01)
\end{itemize}
\noindent In both examples, participants provided a specific idea they wanted to write about but relied on ChatGPT to generate and/or strengthen the paragraph.

Another interesting pattern we observed was the participants providing an incomplete sentence or paragraph and asking ChatGPT to complete it (TR02). The following query exemplifies this pattern well.
\begin{itemize}
    \item P04: \textit{I need my first intro sentence and I've got a start: ``Another way automation, especially AI, is a good thing to help us expose our own inefficiencies is '' But I don't know how to finish it} (TR02)
    \item P15: \textit{As human beings, we strive to continue progressing toward a better future. Automation has been at the forefront of} (TR02)
\end{itemize}

\noindent In P15's case, a clear prompt was missing; the participant expected ChatGPT to complete an unfinished sentence, and the model returned an entire essay continuing from that fragment, although P15 only used one sentence from it. One could argue that the participant did not provide a clear idea of what to write next and that this case should have been categorized under the \textit{All} category. However, since a seed idea was present and the cues from the incomplete sentence guided the direction of the essay, we interpreted it as the participant’s rough idea of what the generated text should be.

In general, Translating queries were not common, suggesting that when participants wanted to generate text, they typically did not separate the task into two distinct queries, Planning and Translating. Alternatively, this may indicate that when participants had an idea of what to write---either on their own or derived from a Planning query---they were generally willing to write independently.

\subsubsection{\textbf{Reviewing}}
For the Reviewing type of ChatGPT queries, 34 out of 77 participants submitted at least one. These queries typically included the written essay as part of the input. The most common code was \textit{Proofreading} (RE01), where participants asked ChatGPT to refine, polish, or proofread their work. Examples of Proofreading queries are presented below.
\begin{table}[h!t]
    \centering
    \caption{\textbf{Reviewing} Query Types}
    \label{tab:reviewing}
    \begin{tabular}{c p{0.42\linewidth} c c}
            \toprule
Code & What did they ask ChatGPT to do? & Count & Participants (\%) \\
        \hline
        RE01 & Proofread & 27 & 16 (20.8\%) \\
        RE02 & Answer spelling/grammar questions & 13 & 10 (13.0\%) \\
        RE03 & Give feedback & 19 & 7 (9.1\%) \\
        RE04 & Shorten text/remove some content & 8 & 7 (9.1\%) \\
        RE05 & Rewrite existing text based on a user's prompt & 11 & 6 (7.8\%) \\
        RE06 & Improve the essay & 8 & 6 (7.8\%) \\
        RE07 & Check if the essay meets the prompt & 4 & 3 (3.9\%) \\
        \bottomrule
    \end{tabular}
    \Description{Top Reviewing query codes ranked by frequency. Our most common code, RE01 - "Proofread",  accounts for 27 instances across 16 (20.8 percent) of participants.}
\end{table}

\begin{itemize}
    \item P03: \textit{Check the essay for inaccurate or unclear statements} (RE01)
    \item P27: \textit{Review this essay and make recommendations for grammar, spelling, punctuation and clarity of thought} (RE01)
\end{itemize}


\noindent In addition to proofreading, participants sometimes asked about specific word spellings or grammar, for example, ``\textit{how to spell sophisticated [sic]} (P60)'' or ``\textit{is `extremely faster' correct?} (P63).''

Another common type of Reviewing was asking ChatGPT to provide feedback and evaluate the essay (RE03, RE07). Once an essay was written, some participants asked ChatGPT to give feedback on it or even to grade it as if it were an assignment. Seven participants requested general feedback on the content of their essays, with examples presented below:
\begin{itemize}
    \item P04: \textit{Here's our first paragraph now expanding on my opinion, thoughts? [a paragraph text]} (RE03)
    \item P06: \textit{Now as the teacher of the class, grade this essay based on if the essay meets this: 
    state your own perspective on the issue, and analyze the relationship between your perspective and at least one other perspective on the issue.
    Grade the essay out of 100.} (RE03)
\end{itemize}


\noindent This type of query corresponds to evaluation, which, along with revision, is one of the two components of the Reviewing process in \citeauthor{flower_cognitive_1981}’ model. However, it should be noted that participants made little effort to specify how ChatGPT should evaluate the essay (e.g., in terms of content, organization, or clarity) beyond simply including the writing prompt. Rather, a couple of evaluations ask ChatGPT to explicitly grade the essay (e.g. \textit{``Grade the essay out of 100''}, P06), which we can see as their goal-oriented attitude. 



Lastly, several participants asked ChatGPT to \textit{revise or rewrite a paragraph} in some way. We did not categorize these queries as Translating, since participants provided existing text and did not explicitly request changes to the underlying ideas; such cases were categorized as All queries. Instead, these requests focused on altering the writing style (RE05), shortening or removing text (RE04), or improving overall quality (RE06).
\begin{itemize}
    \item P26: \textit{I think your revisions are good, but I think my original paragraph had more of a human touch to it. Do you think you can add that human touch back in to your revision?} (RE05)
    \item P65: \textit{how can the following essay be improved [essay pasted]} (RE06)
    \item P65: \textit{Make this sound more professional: [a paragraph pasted]} (RE06)
\end{itemize}

\noindent In these examples, the participants delegated the revision task to ChatGPT. However, they at least evaluated the essay themselves and determined that the current version was insufficient, prompting them to seek improvements in specific ways.




    \subsubsection{\textbf{All}}
The \textit{All} category was used when the participants made a query that involved assistance spanning all three processes of writing. More than 50\% of participants used at least one All query, with a total of 137 such queries, an average of 3.3 per person among those who used them. All queries were the most common query type overall.
\begin{table}[h!t]
    \centering
    \caption{\textbf{All} Query Types}
    \label{tab:all}
    \begin{tabular}{c p{0.42\linewidth} c c}
            \toprule
Code & What did they ask ChatGPT to do? & Count & Participants (\%) \\
        \hline
        AL01 & Generate an essay entirely & 20 & 20 (26.0\%) \\
        AL02 & Write conclusion & 13 & 13 (16.9\%) \\
        AL03 & Generate an alternative essay with some feedback & 29 & 12 (15.6\%) \\
        AL04 & Generate a portion of an essay given a high-level idea & 23 & 11 (14.3\%) \\
        AL05 & Generate the entire essay given a high-level idea & 15 & 8 (10.4\%) \\
        AL06 & Shorten/Lengthen the generated text from the response & 17 & 7 (9.1\%) \\
        AL07 & Write introduction & 6 & 4 (5.2\%) \\
        \bottomrule
    \end{tabular}
    \Description{Top All query codes ranked by frequency. Our most common code, AL01 - "Generate an essay entirely",  accounts for 20 instances across 20 (26.0 percent) of participants.}
\end{table}

20 participants simply asked ChatGPT to \textit{generate an entire essay} (AL01), making this the most frequently used code among all query types. In most cases, they copied and pasted the essay prompt with little additional input reflecting their own thoughts. This type of query was often the first one issued to ChatGPT, meaning that many participants began the writing process with a draft already generated by the system.

Some participants provided additional context on what they wanted in the essay, often adding a few sentences to express their opinion on the issue (AL05). A few examples of a participant adding their opinion are shown below:
\begin{itemize}
    \item P77: \textit{Pretend you are a college student who needs to write an essay on your perspective on the use of machines and artificial intelligence in society. You believe that progress is generally a good thing, but we should be wary of the dangers of overreliance. This is the prompt: [Pasted the writing prompt that contains three perspectives]} (AL05)
    \item P24: \textit{Issue is about Automation and how humans can being [sic] replaced with machinery. Write an essay about how it could affect humanity in a dystopian view} (AL05)
\end{itemize}

\noindent In these examples, the choice of one perspective among the three provided in the prompt was included as part of the query (AL04). This pattern of giving high-level direction was often used for the generation of smaller portions of the essay (e.g., a few paragraphs). A similar pattern was observed in other common queries, such as requesting an introduction (AL07) or a conclusion (AL02). When given a body of text, whether written by themselves or by ChatGPT, participants frequently asked the system to generate a missing paragraph without providing clear instructions.

Another common pattern was when a participant read what ChatGPT generated and then asked it to \textit{rewrite} the text after providing some feedback. This process involved evaluating the generated essay and identifying areas for revision. Sometimes, the requested revision focused on length (AL06), but more often, participants provided specific feedback for ChatGPT to incorporate (AL03). Here are some examples:
\begin{itemize}
    \item P20: \textit{dont include self driving cars, give another good example that involves a readily applicaiton [sic] used by people daily. like Siri by Apple} (AL03)
    \item P05: \textit{rewrite in college level} (AL03)
    \item P61: \textit{make it souund intelligvent [sic]. and make it long} (AL03, AL06)
\end{itemize}


\noindent As seen in the examples, we did not categorize these as Translating or Reviewing queries because (1) the provided ideas were not concrete enough to guide what ChatGPT would generate, and (2) the essay they asked to revise had also been generated by ChatGPT.
Overall, the result showed that when students were in a situation where their ChatGPT usage was not regulated, they used it to conveniently generate the entire essay.

\subsubsection{\textbf{Vibe Writing: Writing with the agent}}
\label{sec:vibe}


%
 
One notable pattern we observed was that some participants used ChatGPT to generate most or all of the text for their essay while steering the output with higher-level feedback and prompts. 

For example, participants could write an entire essay through ChatGPT interactions without typing a single word in the editor, simply by pasting what the system generated; 15 participants (19.5\%) submitted essays that were 100\% written by ChatGPT. 
This pattern parallels Vibe Coding~\cite{sarkar2025vibe}, where a programmer collaborates with AI to produce code aligned with their intended specification. We refer to this mode as \textit{Vibe Writing} and present it in this section.

%
 
Vibe writing does not necessarily mean it only contains queries in All categories; it can involve some queries that are classified as Planning, Translating, or Reviewing. 

For example, 
text generated from Planning queries---intended to provide ideas---was sometimes used directly in the final essay, as they could copy and paste text from the generated response. In fact, participants copied and pasted responses from Planning queries a total of 14 times, across seven individuals. This behavior suggests that even when writers asked ChatGPT for ideas or information, the generated text was often copied and pasted directly into their final essays, possibly diverging from their original intention.

\begin{table*}[h!t] \centering 
  \caption{Query count and query-type distribution (\%) per group. We bold the primary query type when its average proportion is greater than, or approximately equal to, half of the total.} 
  \label{tab:cluster} 
\begin{tabular}{@{\extracolsep{5pt}} ccccccc} 
\\[-1.8ex]\hline 
\hline \\[-1.8ex] 
 & \multicolumn{2}{c}{Group Statistics} & \multicolumn{4}{c}{Proportion of Query Distribution\%} \\  
\cmidrule(lr){2-3}
\cmidrule(lr){4-7} 
Group & Size & Query Count (SD) & Planning & Translating & Reviewing & All \\  
\hline \\[-1.8ex] 
No Query & 6 & 0 (0) & 0 & 0 & 0 & 0 \\ 
Planning & 16 & 2.62 (1.96) & \textbf{93.0\%} & 2.1\% & 1.3\% & 3.6\% \\ 
Translating & 5 & 4.8 (2.17) & 5.0\% & \textbf{48.5\%} & 29.5\% & 17.0\% \\ 
Reviewing & 9 & 3.33 (3.16) & 4.4\% & 3.7\% & \textbf{88.1\%} & 3.7\% \\ 
All & 25 & 5.28 (6.09) & 0.9\% & 0.4\% & 3.5\% & \textbf{95.1\%} \\ 
Mixed & 16 & 7.44 (6.3) & 33.2\% & 3.1\% & 37.6\% & 26.1\% \\ 
\hline \\[-1.8ex] 
\end{tabular}
\Description{Group characteristics showing variation in query patterns. Table shows the size and average query code for each group. Table also shows proportion of the query distribution.}
\end{table*} 
\begin{figure*}[ht!]
    \centering
    \begin{subfigure}[b]{0.49\textwidth}
        \centering
        \includegraphics[width=\textwidth]{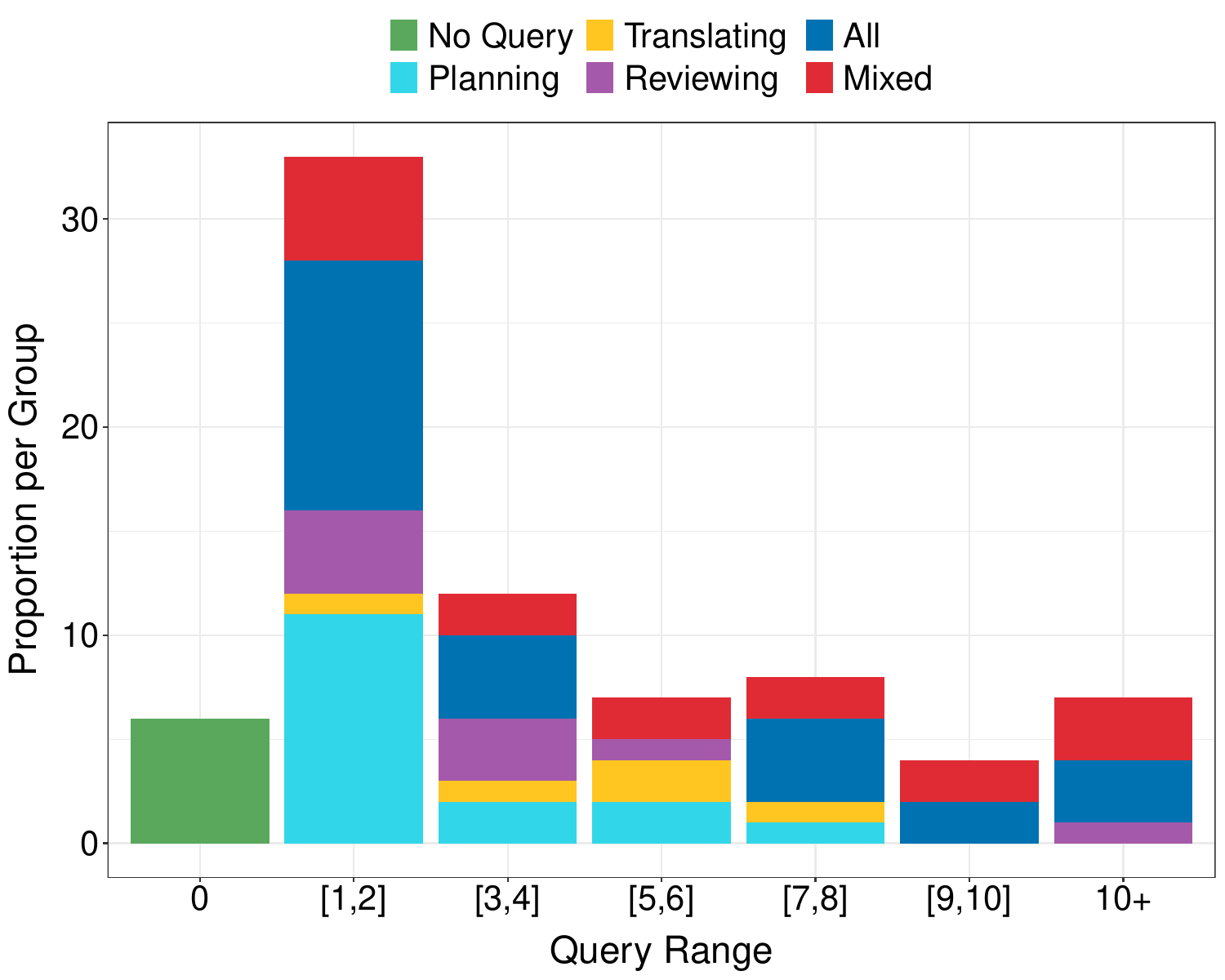} 
        \caption{Histogram of the number of queries asked (in bins of 2)}
        \Description{Stacked histogram of queries asked showing right-skewed query distribution. Most participants concentrated in low ([1.,2] or [3,4]) query bins, with varied group distribution.}
        \label{fig:hist} 
    \end{subfigure}
     \hfill 
    \begin{subfigure}[b]{0.49\textwidth} 
        \centering
        \includegraphics[width=\textwidth]{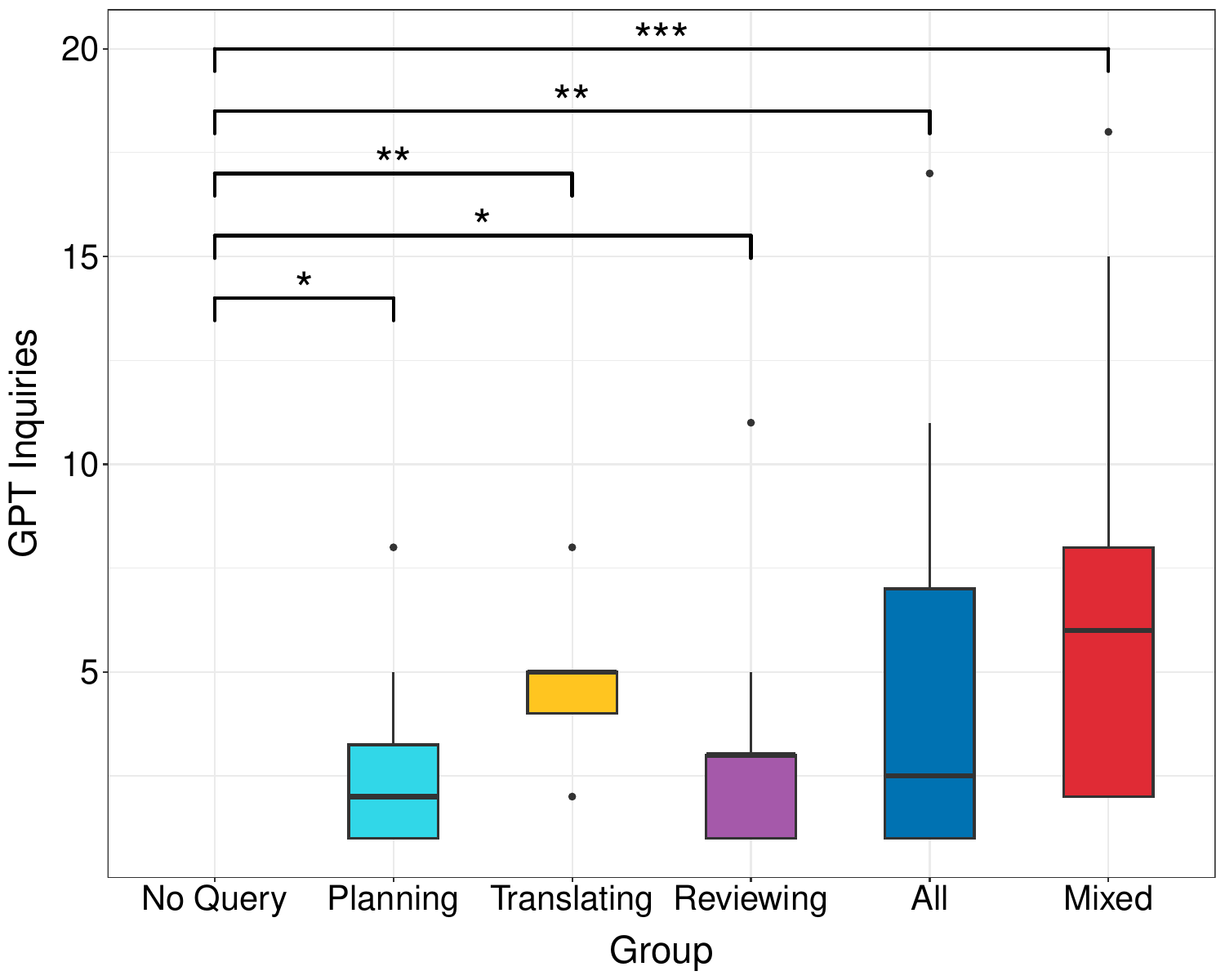}
        \caption{Box plot showing query counts per group}
        \label{fig:groupinquiry} 
        \Description{Box plots showing query count variation across groups. Clusters vary in position between 0-20+ queries.}
    \end{subfigure}
    \caption{Query distribution per group}
    \Description{Stacked bar chart and Box plot showing the distributions of query across groups. The leftmost graph shows the stacked bar chart of the different group types in the query bins. The rightmost box plot shows significant differences in query count between groups.}
    \label{fig:inquiry_dist} 
\end{figure*}
Using a Reviewing query also did not necessarily mean that participants revised their own writing with ChatGPT. A preceding query could have generated text based on an All-type query. For example, P74's essay is a clear example of Vibe Writing mode: the participant produced a 641-word essay, 87.1\% of which was written by ChatGPT across 11 queries. The following two examples illustrate the types of queries that P74 used.

\begin{itemize}
        \item P74: \textit{First, let’s say ``a'' perspective instead of ``the'' perspective. Second, I don’t see any lines that hint at the unique perspective of the author, it’s just an overview of the other two perspectives. Let’s add something like:  
        ``A lot of labor in today’s world involves tasks that don’t necessarily require human levels of intelligence to complete, because the end product is well defined. A utilitarian perspective on automation acknowledges this....''  
        Also you could be a bit more detailed in your ideas—just act like you’re a 1900s sci-fi writer or someone who would have thought about this enough.}  
        \item P74: \textit{That conclusion is a bit too whimsical! Channel your inner science journal.}  
    \end{itemize}
\noindent 
One might argue that using Planning, Translating, and Reviewing queries is seemingly less problematic than relying on All queries. However, in practice, participants often interacted with ChatGPT to generate ideas, produce text, and revise it by prompting only, effectively authoring an essay in the mode of Vibe Writing.


The result shows the complexity of understanding the impact of GenAI on students’ learning; it is necessary to look beyond the types of queries they made. Based on our findings, multiple components need to be considered: how GenAI responded to a query, how a student integrated the generated text into their writing, what kinds of text were used in the query and where that text originated, and the overall history of queries, responses, and copied text.



\subsubsection{\textbf{Clustering Participants by ChatGPT Usage Patterns}}
\label{sec:analysis34}

 For further analysis of each research question, we clustered participants based on the similarity of their usage patterns. We employed K-means clustering to identify underlying patterns in student use of ChatGPT by grouping participants according to the distribution of their query types. This unsupervised machine learning approach enabled data-driven classification and provided deeper insight into distinct usage behaviors across groups and how different usage patterns affect their experiences (e.g., Perceived Ownership in RQ4).

 Each participant was represented by a 4-dimensional feature vector, where each component corresponded to the percentage of P, T, R, and A query types out of all queries they submitted. For example, if a participant asked one planning query and three reviewing queries, their feature vector would be $[0.25, 0, 0.75, 0]$. Participants who did not ask any questions were represented with a zero vector.

 Using these feature vectors, we grouped participants by their primary ChatGPT usage. Based on the sum of squared error (SSE) scores (shown in Appendix \autoref{fig:SSE}), we applied the elbow method~\cite{thorndike_who_1953} and determined that $K = 6$ was appropriate. These clusters were visually inspected based on the query type distribution and named according to defining characteristics (shown in Appendix \autoref{fig:cluster_prop}): 

\begin{itemize}[left=0.3 cm]
    \item \textbf{Group No Query (N)}:  Participants who did not use ChatGPT for the task ($n=6$).
    \item \textbf{Group Planning (P)}:  Participants whose ChatGPT queries were primarily focused on Planning activities ($n=16$).
    \item \textbf{Group Translating (T)}:  Participants whose ChatGPT queries were primarily focused on Translating activities ($n=5$). This group's participants also contain very low Planning queries.
    \item \textbf{Group Reviewing (R)}:  Participants whose ChatGPT queries were primarily focused on Reviewing activities ($n=9$).
    \item \textbf{Group All (A)}:  Participants whose ChatGPT queries were primarily focused on All activities ($n=25$).
    \item \textbf{Group Mixed (M)}:  Participants with mixed behaviors, whose ChatGPT queries were distributed across multiple categories ($n=16$). The participants in this group used few Translating queries.
\end{itemize}
 \noindent The average distribution of each query type per group is presented in \autoref{tab:cluster}.

 Using these groups, we tested whether there were significant differences in the metrics examined in RQ3 and RQ4 based on group membership. Because none of the metrics met the assumption of normality (e.g., average values of ordinal outcomes), we used the Kruskal–Wallis test. For post hoc analysis, we used Dunn tests with Bonferroni correction to identify significant groupings.

\begin{figure*}[ht!]
    \centering
    \begin{subfigure}[b]{0.49\textwidth}
        \centering
        \includegraphics[width=\textwidth]{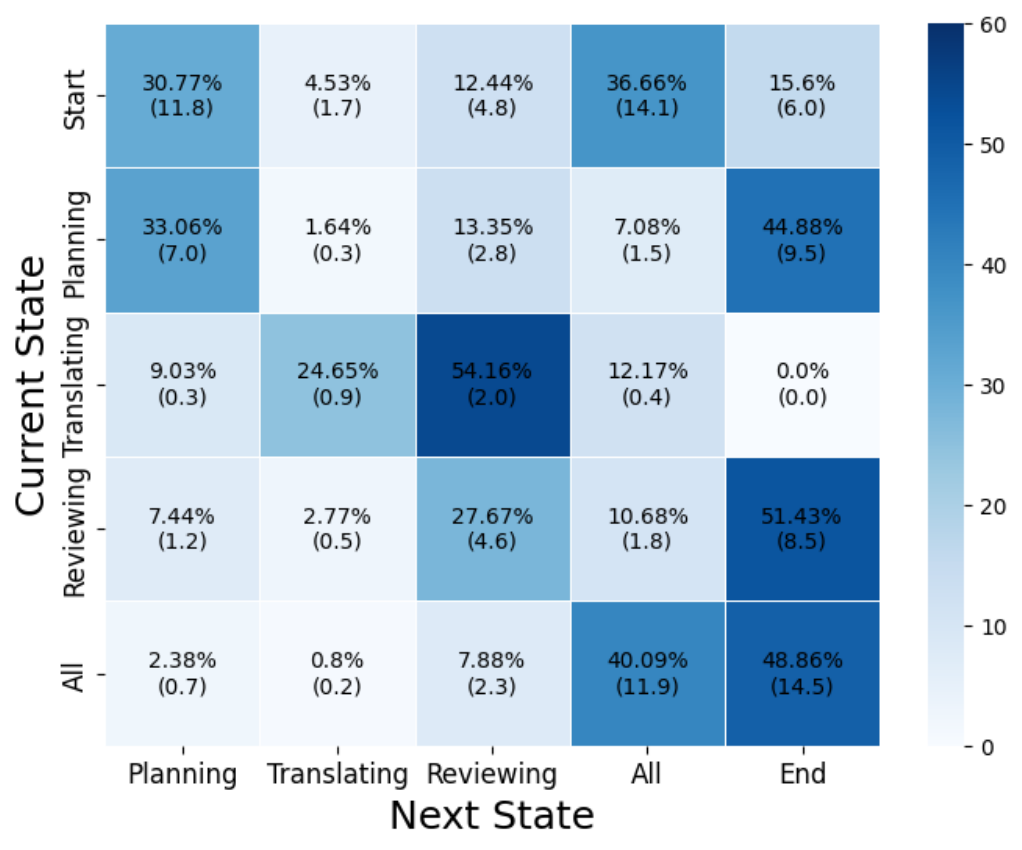} 
        \caption{Transition matrix for the queries. Each cell shows the probability of transitioning to a state (column) given the current state (row). The number in parentheses indicates the number of times the transition was observed, normalized to the total number of queries made by each participant.} 
        \Description{Transition matrix with heatmap showing query switching patterns. Both self and cross-category patterns are prominent in the graph with Planning, Reviewing, All transitioning to end accounting for the major transitions.}
        \label{fig:trans} 
    \end{subfigure}
     \hfill 
    \begin{subfigure}[b]{0.49\textwidth} 
        \centering
        \includegraphics[width=\textwidth]{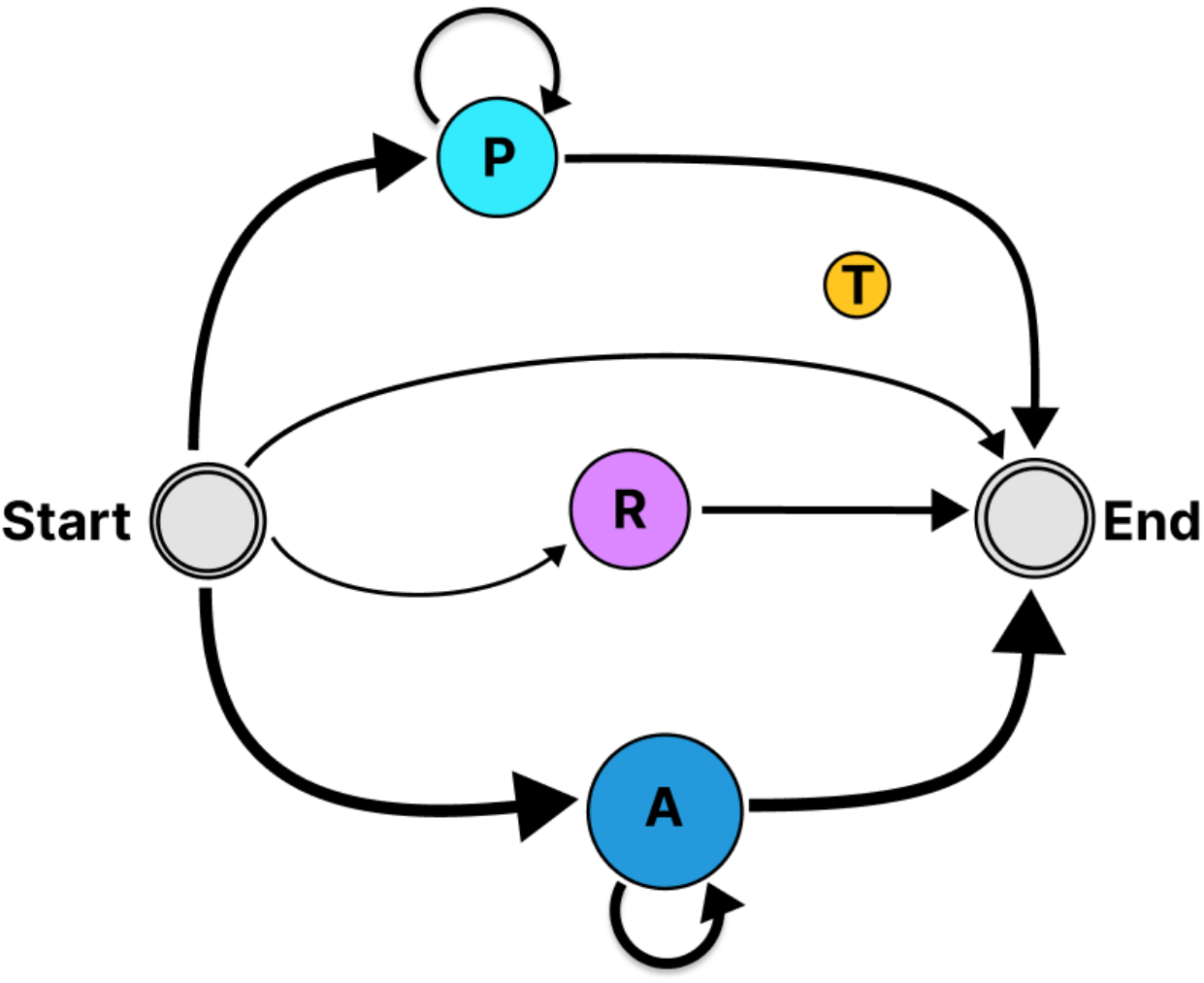}
        \caption{State diagram representing the transition matrix. Node size reflects the total number of queries, and edge width is proportional to the number of transitions. This diagram displays only the 9 most frequent edges where the number of transitions accounts for 80\% of the entire transitions.}
        \Description{State diagram with node and edge sizing showing pathway frequency. Linear patterns are prominent between nodes.}
        \label{fig:state} 
    \end{subfigure}
    \Description{Transition Matrix and State Diagram showing the most common query transitions used by participants}
    \caption{Transition matrix and state diagram for ChatGPT queries. The leftmost graph is the tranistion matrix heatmap, while the right shows the state diagram.}
    \label{fig:state_matrix} 
\end{figure*}

\subsubsection{\textbf{Interaction Patterns and Writing Stage Transitions}}

We examined how many times each participant asked questions to ChatGPT to show the distribution of query counts, as shown in \autoref{fig:hist}.
Overall, nearly half of the participants (39 out of 77) used ChatGPT less than or equal to twice; six participants, classified as Group N, never used it at all.
Meanwhile, some participants used ChatGPT multiple times, with one participant submitting 27 queries, the highest count observed. We analyzed the number of queries submitted by each group using a box plot (\autoref{fig:groupinquiry}), but we did not find statistically significant differences between our ChatGPT-using groups except for Group No Query, which differed from all other groups. 
Based on the proportion of GPT-generated words, participants who wrote in Vibe Writing mode are more likely to belong to Group A and Group M, given the significantly higher shares of GPT-produced text in their final essays.


We analyzed the sequence of query types by constructing a transition matrix and visualizing the typical order of interactions through a state diagram. A transition probability in the matrix (\autoref{fig:trans}) represents the likelihood of a participant asking a particular type of query given their current state. We also report the normalized counts of each transition, shown in parentheses in \autoref{fig:trans}. To account for individual differences in the total number of queries and to avoid over-weighting participants who asked many queries, transition counts were normalized by dividing each by the total number of transitions per participant. Using these normalized counts, we generated a state diagram in \autoref{fig:state} that highlights transition paths representing 80\% of all transitions, thereby revealing the common sequence participants followed.
\begin{table*}[t!] \centering 
  \caption{Generalized Linear Model for Total Query and Query Counts by Code} 
  \label{tab:glm}
\begin{tabular}{@{\extracolsep{5pt}}lccccc} 
\toprule 
 & Total Queries & Planning Queries & Translating Queries & Reviewing Queries & All Queries \\ 
 \hline \\[0.3ex] 
 Self-Efficacy & \textbf{$-$0.258$^{***}$} & $-$0.204 & \textbf{$-$0.949$^{**}$} & \textbf{$-$0.588$^{***}$} & $-$0.016 \\ 
  & (0.075) & (0.147) & (0.307) & (0.154) & (0.114) \\[1.3ex]  
 TAM PU & 0.048 & 0.122 & $-$0.384 & \textbf{$-$0.205$^{*}$} & \textbf{0.282$^{**}$} \\ 
  & (0.062) & (0.131) & (0.208) & (0.103) & (0.106) \\[1.3ex]   
 TAM PEOU & $-$0.087 & 0.007 & $-$0.025 & 0.051 & \textbf{$-$0.277$^{**}$} \\ 
  & (0.066) & (0.141) & (0.215) & (0.119) & (0.103) \\[1.3ex]   
 Gender (Male) & \textbf{0.308$^{*}$} & 0.361 & 0.064 & \textbf{0.712$^{**}$} & 0.030 \\ 
  & (0.125) & (0.252) & (0.455) & (0.270) & (0.185) \\[1.3ex]   
 Age & \textbf{$-$0.038$^{**}$} & $-$0.006 & $-$0.041 & $-$0.002 & \textbf{$-$0.127$^{**}$} \\ 
  & (0.013) & (0.021) & (0.044) & (0.021) & (0.040) \\[1.3ex]   
 Race (White) & \textbf{0.373$^{**}$} & 0.191 & 0.351 & \textbf{0.946$^{***}$} & 0.199 \\ 
  & (0.117) & (0.234) & (0.435) & (0.246) & (0.178) \\[1.3ex]   
 Constant & \textbf{3.534$^{***}$} & 0.262 & \textbf{6.679$^{**}$} & \textbf{3.064$^{**}$} & \textbf{3.359$^{**}$} \\ 
  & (0.581) & (1.064) & (2.546) & (1.154) & (1.149) \\ 
  & & & & & \\ 
\bottomrule 
\\[.8ex] 
\textit{Note:}  & \multicolumn{5}{r}{$^{*}$p$<$0.05; $^{**}$p$<$0.01; $^{***}$p$<$0.001} \\
\end{tabular} 
\Description{Generalized linear model results showing predictor relationships. Multiple significant predictors for ChatGPT usage.}
\end{table*} 

The state diagram reveals four common paths among participants. The first is $ \text{Start} \rightarrow \text{Planning}(+) \rightarrow \text{End} $, where the plus sign ($+$) indicates that the state can be repeated multiple times. Participants on this path asked only Planning questions before completing their essays. The second path is $ \text{Start} \rightarrow \text{Reviewing} \rightarrow \text{End} $, where participants used ChatGPT solely to review their writing before finishing the essay. The third path, $ \text{Start} \rightarrow \text{All}(+) \rightarrow \text{End} $, represents participants who asked ChatGPT to generate content that could be directly incorporated into their essays, categorized as \textit{All}. Finally, some participants did not use ChatGPT at all ($ \text{Start} \rightarrow \text{End} $). These four paths correspond to four of the six identified groups: Group P, Group R, Group A, and Group N, respectively.

Our analysis of ChatGPT queries revealed several key results: the common types of queries and illustrative examples categorized under the framework of Planning, Translating, Reviewing, and All; the identification of a distinctive ``Vibe Writing'' mode of interaction; and the common sequences in which participants used ChatGPT during the essay writing task.

\subsection{RQ2: The Relationship Between ChatGPT Usage and Participant Perception Towards Their Writing Efficacy and Technology Acceptance}

 For RQ2, we aimed to discover if certain student characteristics might influence their ChatGPT usage patterns. 
For student characteristics, we collected demographic information along with two constructs we measured before the study: the Self-Efficacy in Writing Scale (SEWS) and the Technology Acceptance Model (TAM).

We ran a generalized linear model (GLM) to analyze the relationships between multiple predictor variables and different query types simultaneously.
 Table~\ref{tab:glm} presents the full results, including the estimated coefficients and their significance levels. The sign of each coefficient indicates the direction of the relationship, with positive values reflecting a positive association and negative values indicating a negative association.

SEWS and ChatGPT usage showed a statistically significant negative relationship between total query counts and Translating and Reviewing queries. This result suggests two notable patterns. First, self-efficacy in writing is a significant predictor of ChatGPT use: participants with lower writing self-efficacy tended to use ChatGPT more frequently. Second, this negative relationship was largely accounted for by Translating and Reviewing queries---both closely tied to the writing stage---while no such effect was observed for ideation-related query types (Planning and All).

In contrast, the Technology Acceptance Model (TAM) did not predict the total number of queries, suggesting that participants’ willingness to use ChatGPT was not strongly determined by whether they perceived it as useful (PU) or easy to use (PEOU). However, All queries were predicted in divergent ways by the two TAM components. TAM perceived usefulness (PU) showed a positive relationship, indicating that participants who viewed ChatGPT as more useful were more likely to rely on it to generate full essays. In contrast, TAM perceived ease of use (PEOU) showed a negative relationship, suggesting that participants who found ChatGPT easy to use were less likely to rely on it for All queries, perhaps because they knew various ways to leverage ChatGPT for writing tasks beyond delegating the entire essay. Additionally, TAM PU had a negative relationship with Reviewing queries, indicating that participants who perceived ChatGPT as more useful were less inclined to use it for proofreading or revision tasks.
\begin{figure*}[t!]
    \centering
    \begin{subfigure}[b]{0.49\textwidth} 
        \centering
        \includegraphics[width=\linewidth]{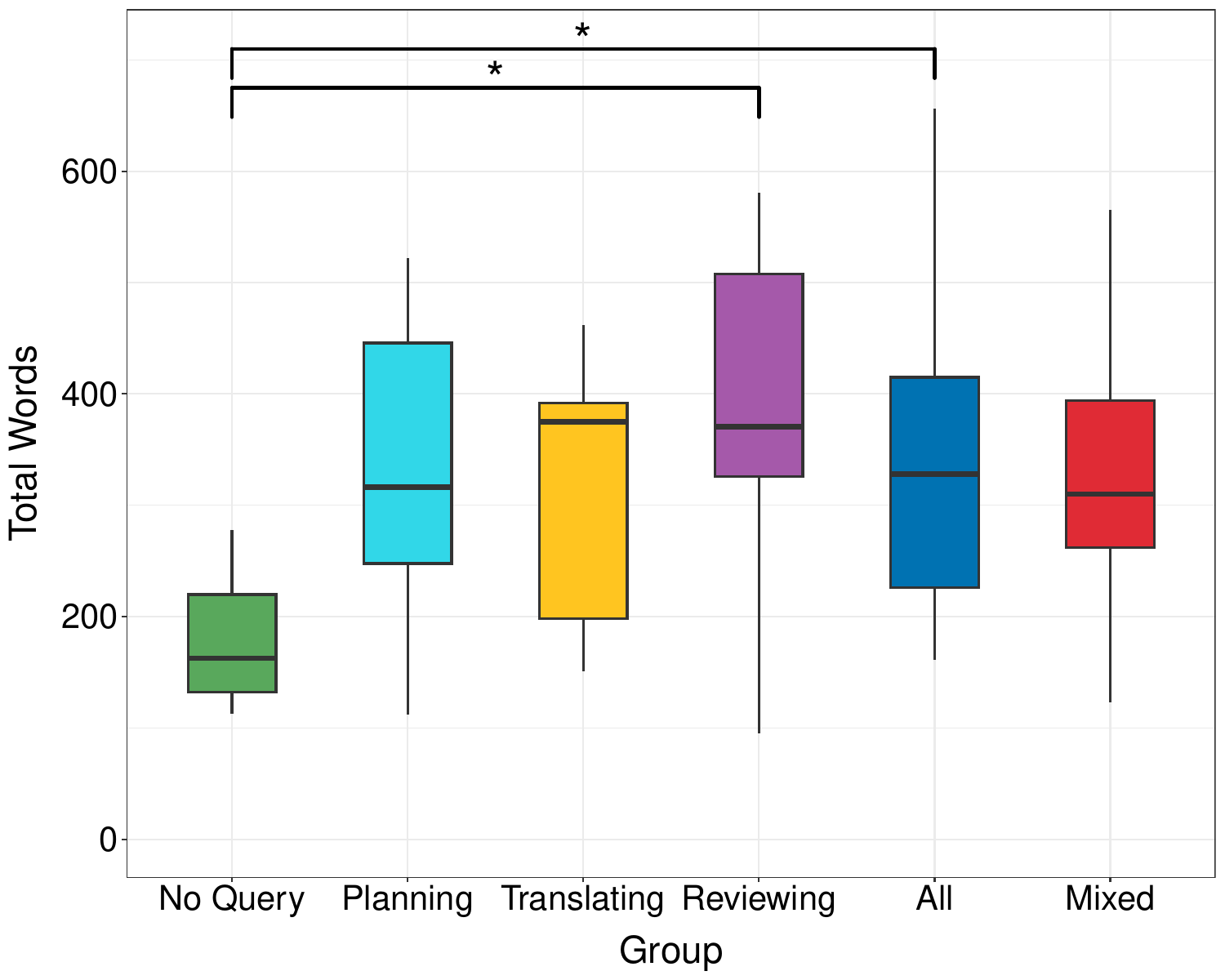}
        \caption{Total words per Cluster}
        \label{fig:total_words} 
        \Description{Box plots showing essay length variation on zero to 600 word scale. Most Groups cluster around 350 to 400 words, with significance for group No Query with Groups Reviewing and All.}
    \end{subfigure}
    \hfill 
    \begin{subfigure}[b]{0.49\textwidth}
       \centering
       \includegraphics[width=\linewidth]{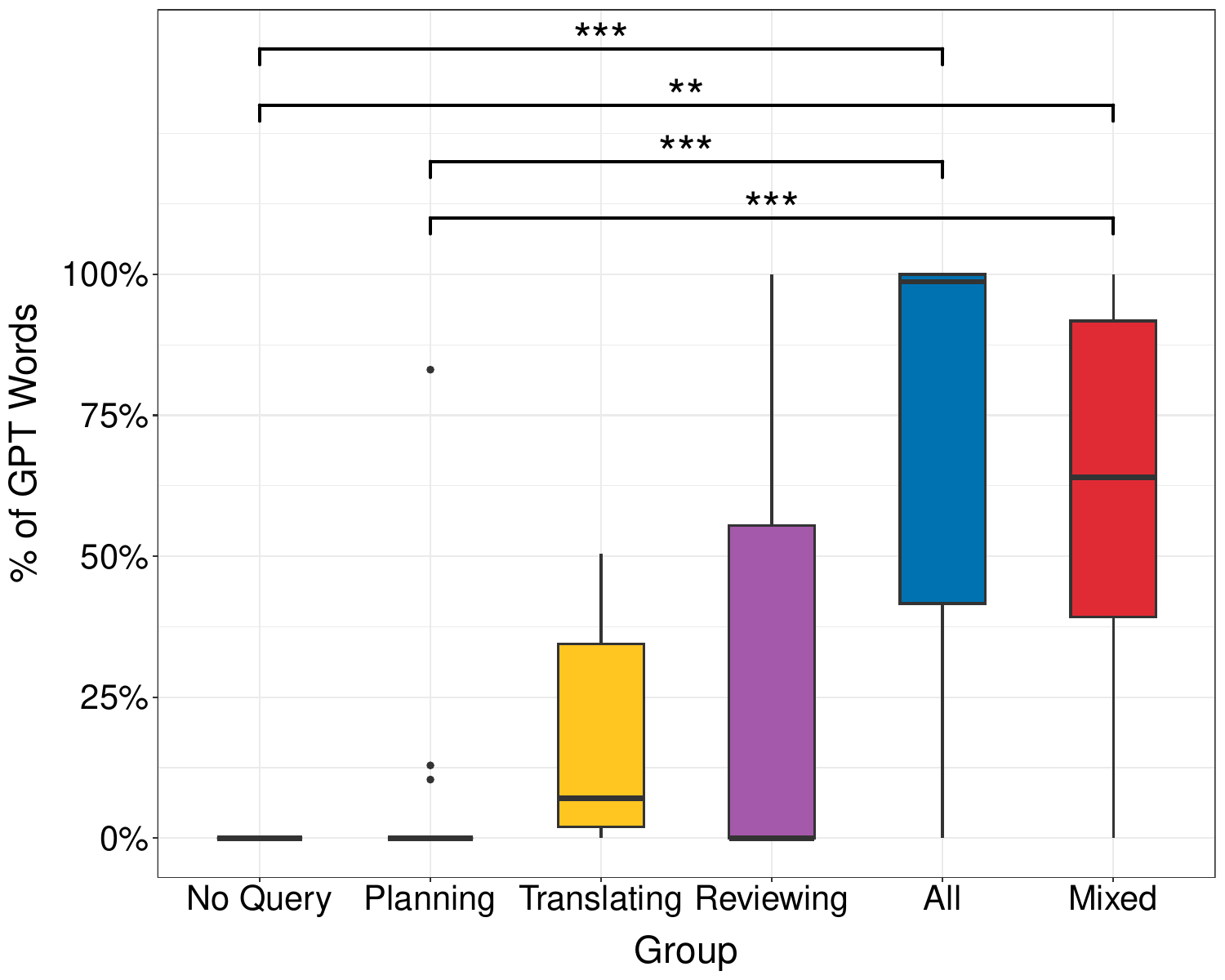}
        \caption{Portion of GPT words in final essay}
        \Description{Box plots showing ChatGPT content proportion across groups. Clear separatio visible with All and Mixed above 50 percent and remaining groups around 0 to 10 percent.}
        \label{fig:gpt_percent} 
    \end{subfigure}
    \caption{Box Plots for Authorship with Post Hoc analysis (* indicates $p < 0.05$)}
    \Description{Box plots showing the significant differences between group type and word distribution. The leftmost graph shows total word, and the rightmost graph shows authorship of the words}
    \label{fig:total_metric}
\end{figure*}
Gender and race were also significant predictors: both Male and White participants submitted a higher number of total ChatGPT queries, as well as more Reviewing queries. Age, by contrast, was a negative predictor in two categories: total number of queries and All queries. Older participants used fewer queries on average, particularly in the All category, suggesting that they were less likely to rely on ChatGPT for full text generation.

\subsection{RQ3: The Relationship Between ChatGPT Usage and Essay Characteristics}
\label{sec:interactiontrace}
 
For RQ3, we wanted to understand how the specific ways students used ChatGPT in their writing process manifested in their final essays.
We analyzed the quantity and quality of the final text, as well as the distribution of human versus AI authorship to assess this relationship. We examined three key characteristics: word count, proportion of the ChatGPT-generated words in the final essay, and readability scores from the Flesch-Kincaid Grade Level and Dale-Chall metrics. Specifically, how these characteristics varied between the groups: \textbf{No Query (N)}, \textbf{Planning (P)}, \textbf{Translating (T)}, \textbf{Reviewing (R)}, \textbf{All (A)}, and \textbf{Mixed (M)}, as described in Section~\ref{sec:analysis34}. The average essay length is shown in \autoref{fig:total_words}. 
In general, participants who used ChatGPT queries wrote longer essays than those who did not use ChatGPT at all. The difference was statistically significant between Group N and Group A and between Group N and Group R.



\begin{figure*}[t!]
    \centering
    \begin{subfigure}[b]{0.49\textwidth} 
        \centering
        \includegraphics[width=\linewidth]{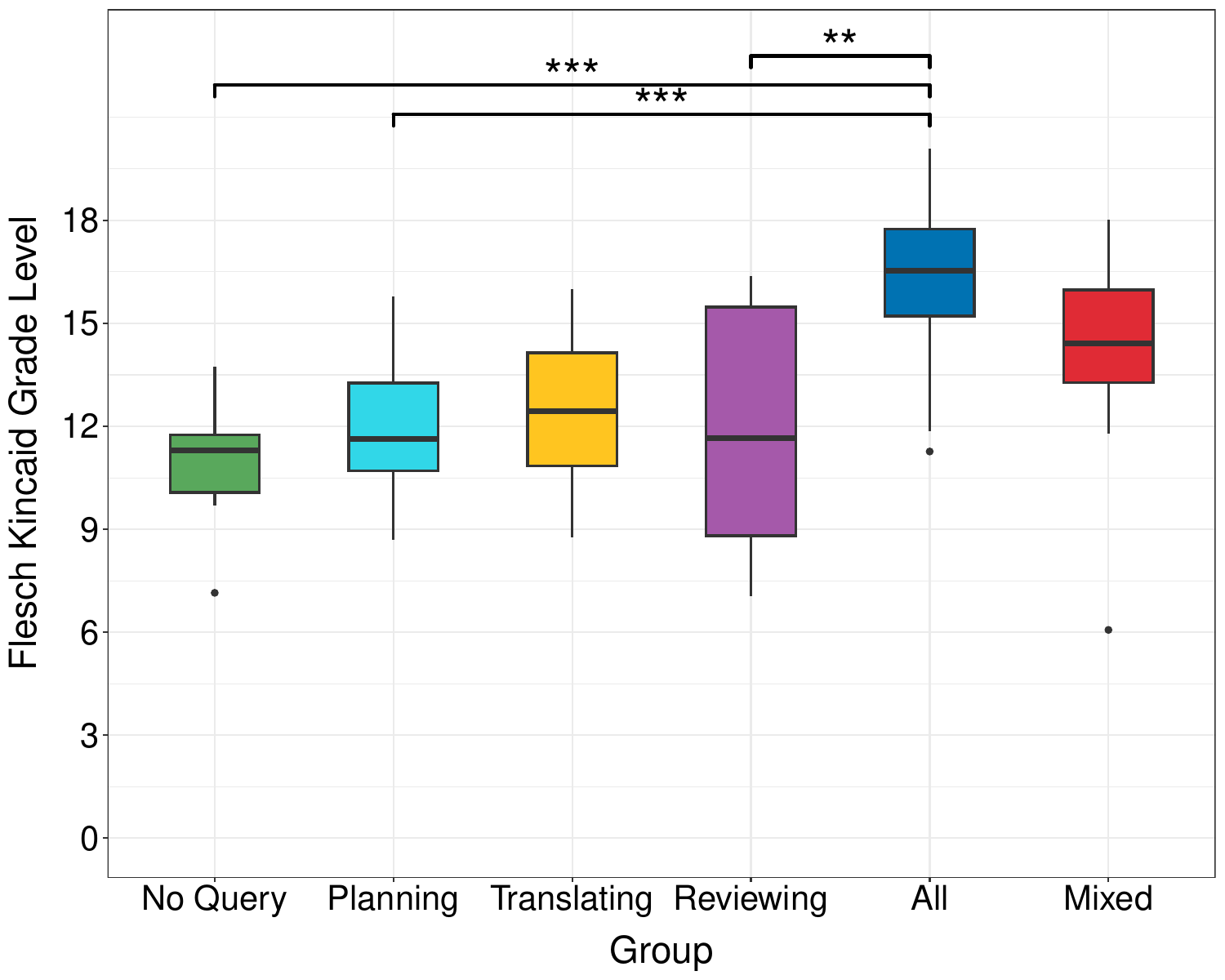}
        \caption{ Flesch-Kincaid Grade Level across Group}
        \label{fig:FK}
        \Description{Box plots showing Flesch-Kincaid Grade Level scores on zero to 18 scale. Multiple groups positioned around grade 12 while All and Mixed cluster around grade 15, showing some significant separation.}
    \end{subfigure}
    \hfill 
    \begin{subfigure}[b]{0.49\textwidth}
         \centering
        \includegraphics[width=\linewidth]{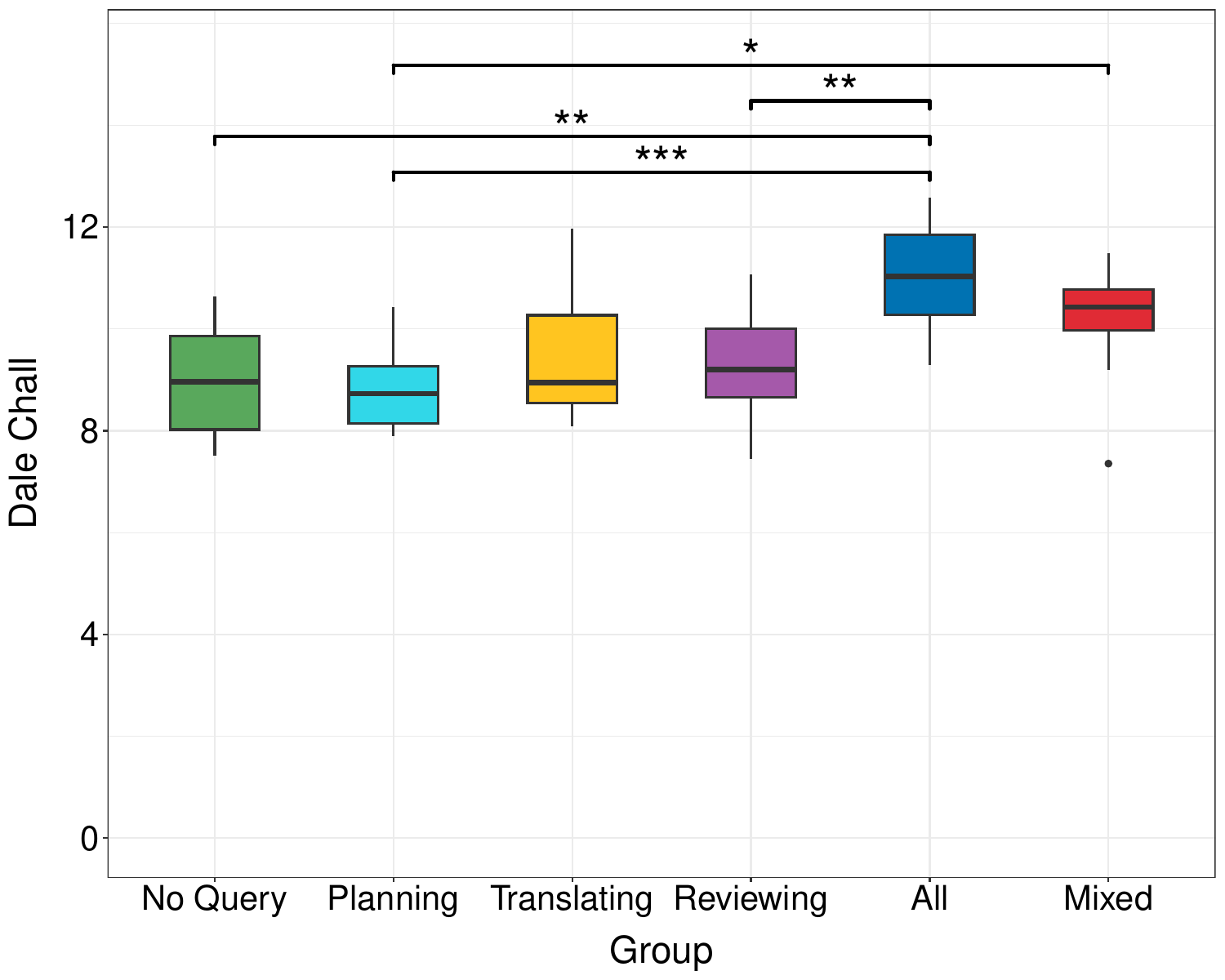}
        \caption{ Dale-Chall Score across Group}
        \label{fig:DC}
        \Description{Box plots showing Dale-Chall readability on zero to 12 scale. Most groups cluster around 9 while other groups positioned around 10 to 11, showing multiple significant differences.}
    \end{subfigure}
    \caption{ Box Plots for Readability metrics with Post Hoc analysis (* indicates $p < 0.05$)}
    \Description{Two box plots showing the different readbility metrics with significant differences. The leftmost graph shows the Flesh-Kincaid Grade level distribution, while the rightmost shows Dale-Chall distributions}
    \label{fig:readability}
\end{figure*}

\subsubsection{Essay Composition}
We tracked both the number of words participants wrote themselves (User Final Word Count) and the number of words pasted from ChatGPT (ChatGPT Final Word Count). In two essays, external words—i.e., words pasted from sources other than ChatGPT or the editor—were added, but they accounted for less than 2\% of the respective essays. We then ran a Kruskal–Wallis test on the proportion of ChatGPT-pasted words to examine whether different groups exhibited different behaviors.
We had multiple significant groups when looking at the percentage of the final essay that was written by ChatGPT (shown in Figure \ref{fig:gpt_percent}). 
Groups N, P, and R resulted in nearly no GPT-pasted words, except for a few outliers in Groups P and R.  
Groups A and M, in contrast, exhibited a wide range of behaviors but a significantly higher proportion of GPT contributing to their final essays. 
A Kruskal–Wallis test revealed significant differences across groups, with post hoc pairwise comparisons indicating significant differences for (N,A), (N,M), (P,A), (P,M), and (R,A) (all $p < .05$).

\begin{figure*}[ht!]
    \centering
    \begin{subfigure}[b]{0.49\textwidth} 
        \centering
        \includegraphics[width=\linewidth]{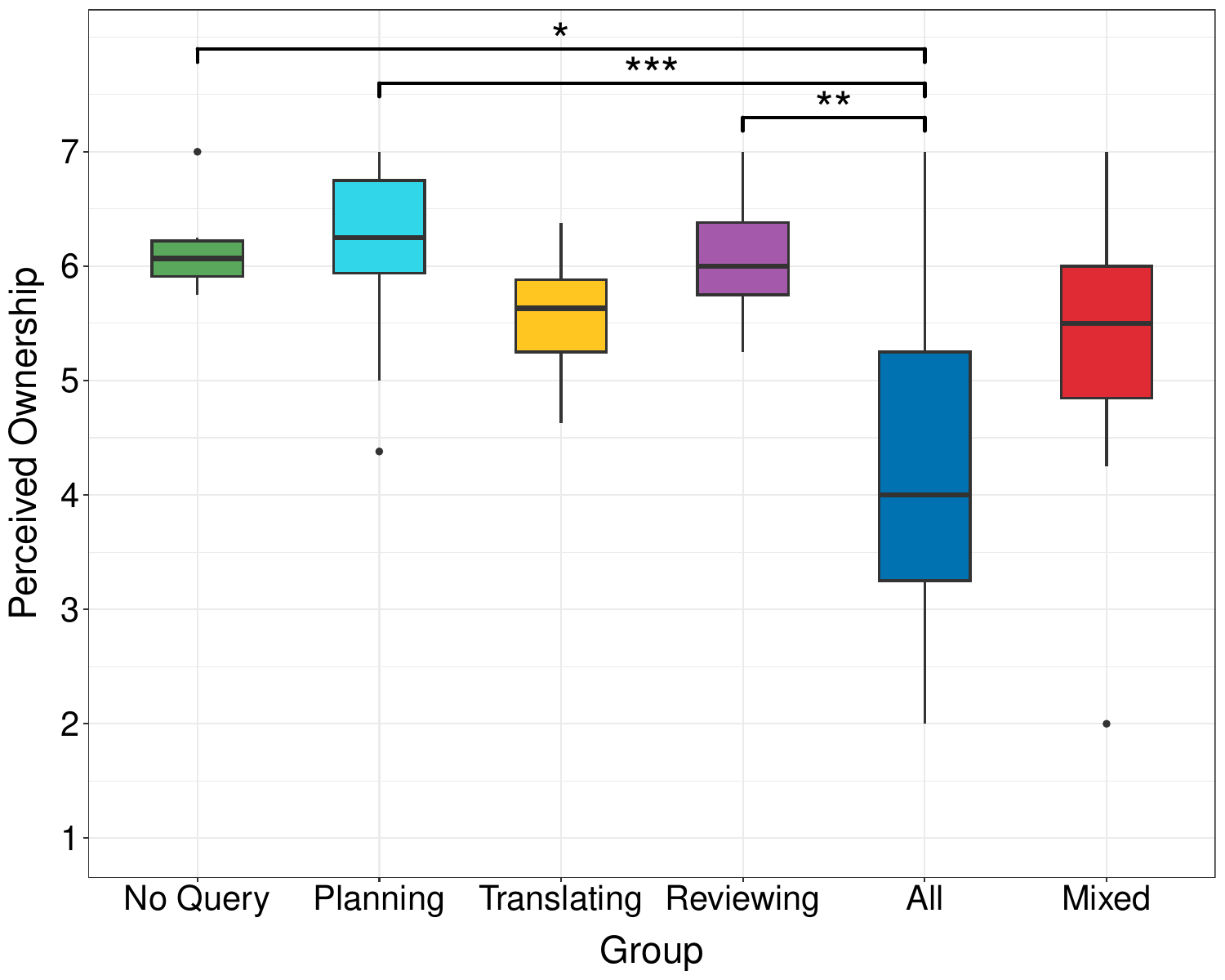}
        \caption{Perceived Ownership across Group}
        \label{fig:PO_Box_Plot}
        \Description{Box plots showing Perceived Ownership ratings on zero to seven scale. Groups show varying spread with significance.}
    \end{subfigure}
    \hfill 
    \begin{subfigure}[b]{0.49\textwidth}
         \centering
    \includegraphics[width=\linewidth]{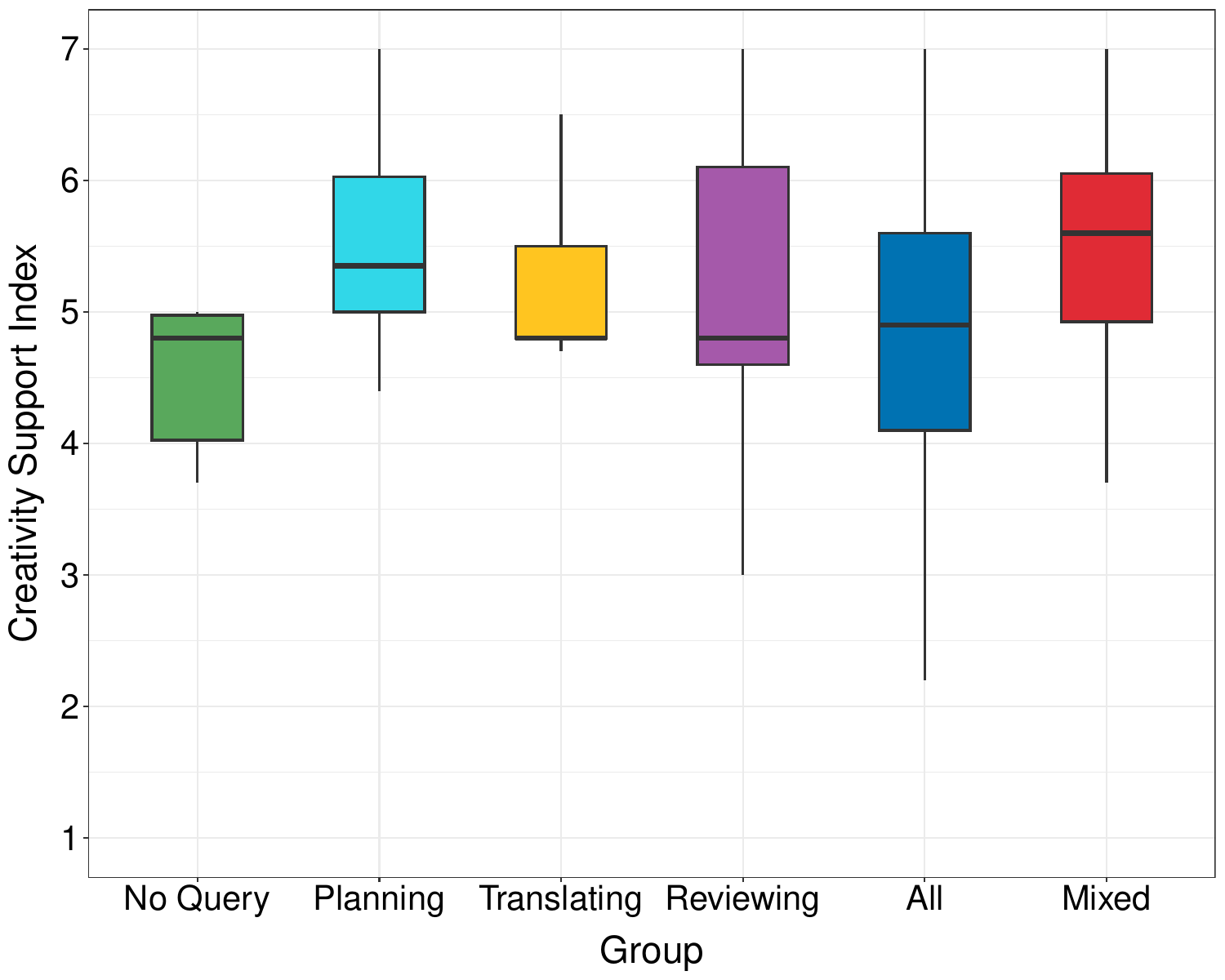}
    \caption{Creativity Support Index across Group}
    \Description{Box plots showing Creativity Support Index on zero to seven scale. Groups positioned similarly around 5 with no significant separation.}
    \label{fig:CSI_Box_Plot}
    \end{subfigure}
    \caption{Box Plots for PO and CSI metrics with Post Hoc analysis (* indicates $p < 0.05$)}
    \Description{Two box plots showing our post-study survey scores with significance. The leftmost graph shows Perceived Ownership while the rightmost shows Creativity Support Index distributions}
    \label{fig:post_study_box}
\end{figure*}

\subsubsection{ Readability scores}

 Using the previously discussed clusters, we ran Kruskal-Wallis testing with post-hoc analysis to look for significant differences (\autoref{fig:readability}). For the Flesch-Kincaid Grade Level scores, we found a $H(5) = 32$, $p < 0.001$. Post-hoc analysis revealed significant differences between Groups (A,N), (A,P), (A,R), with Group A showing significantly higher Grade Level scores. For Dale-Chall, Kruskal-Wallis testing showed a $H(5) = 34$, $p < 0.001$. Post-hoc analysis showed four significant groupings: Group A and Group N, Group A and Group P, Group A and Group R, and Group P with Group M. Both readability metrics showed users in Group A, or users who had ChatGPT write large portions of their essay, produced essays with significantly higher complexity scores than their counterparts; this indicates that ChatGPT generates text with higher complexity. Additionally, the Dale-Chall readability metric reveals a significant difference between Group P and Group M, indicating that the Mixed group had higher complexity scores than the Planning group.


\subsection{RQ4: Participant Perceptions Towards Their Writing Experiences}

 For RQ4, we wanted to understand how students' ChatGPT usage during writing would relate to students' perceptions of their writing.
We asked users to complete a post-study survey focusing on Perceived Ownership and the Creativity Support Index (CSI) to see how their writing experience varied depending on how they used ChatGPT. We ran Kruskal-Wallis tests on this data to see if there were any relationships between these surveys and the writer groupings. For perceived ownership, the group effects were significant, but we had no significance on the total CSI score (shown in Figure \ref{fig:post_study_box}). We then looked at the CSI subcategories and found significance in two subscales. Following these results, we conducted post hoc analysis using a Dunn test with Bonferroni correction similar to that in the previous section.  

For the Perceived Ownership data, the Kruskal–Wallis test showed $H(5) = 27.876$, $p < 0.001$. Post-hoc analysis revealed three statistically significant group differences: between Group N and Group A, between Group P and Group A, and between Group R and Group A (Figure \ref{fig:PO_Box_Plot}). These findings indicate that perceived ownership was significantly lower in the All group compared to the other three groups. This result is unsurprising, as participants in Group A often generated entire essays using ChatGPT and therefore reported lower ownership of the final text. However, the high perceived ownership reported by Groups P and R, comparable to Group N, is noteworthy. From an educational perspective, this may be concerning, as reliance on Planning or Reviewing queries could also influence learning outcomes depending on the objectives of the writing assignment.
Ideally, students' sense of authorship should align with their learning outcomes, as their perceived ownership may contribute to their positive experience of using ChatGPT, potentially without learning gain. This highlights the need for instructors to carefully design writing activities and provide guidance on how AI tools should be used.

\begin{figure*}[ht!]
    \centering
    \begin{subfigure}[b]{0.49\textwidth} 
        \centering
        \includegraphics[width=\linewidth]{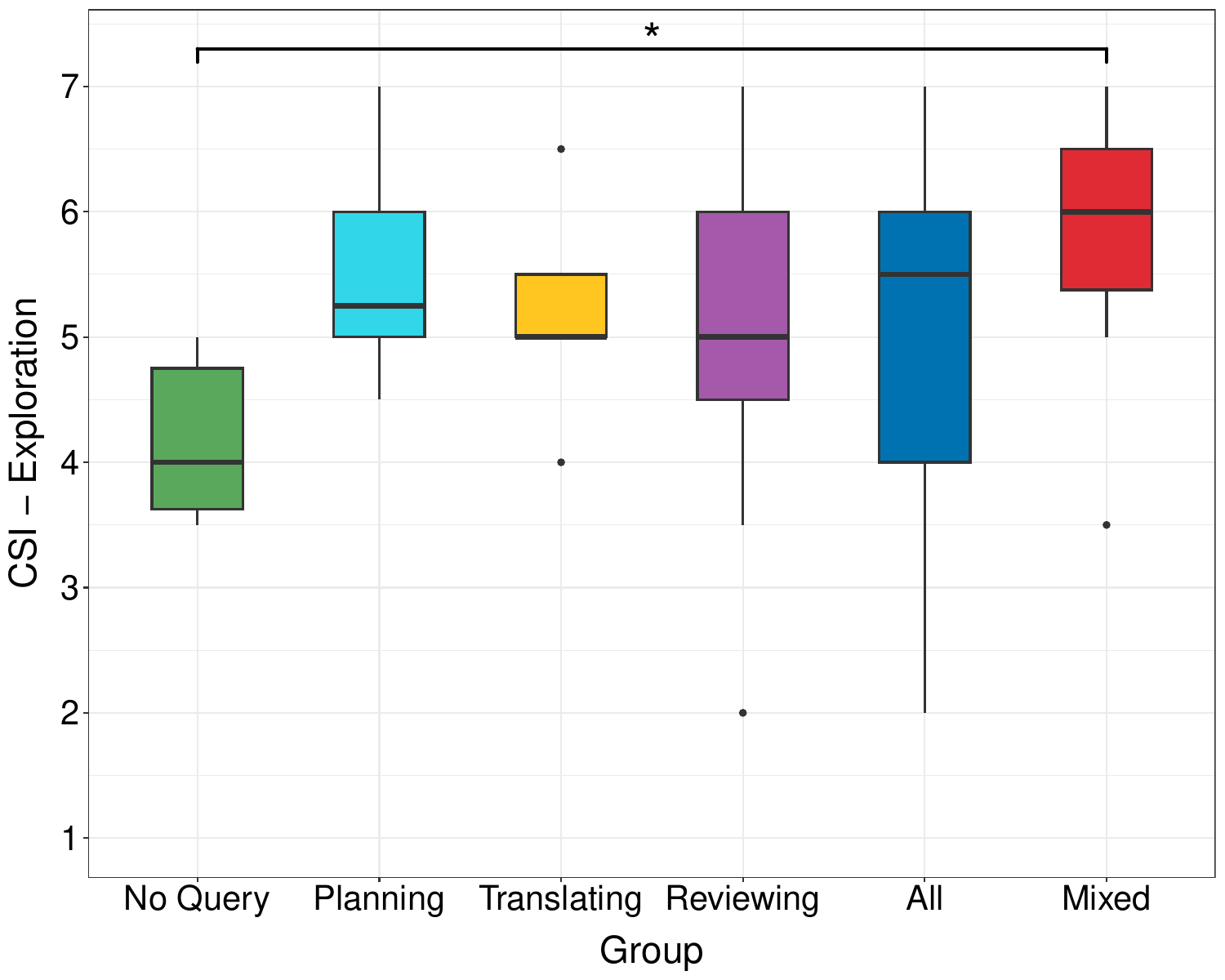}
        \caption{CSI-Exploration across Group}
        \label{fig:csi_exp}
        \Description{Box plots showing Creativity Support Index subscale Exploration scores on zero to seven scale. Most groups cluster around 5, with Mixed and No Query showing significant differences at 4 and 6 respectively.}
    \end{subfigure}
    \hfill 
    \begin{subfigure}[b]{0.49\textwidth}
         \centering
    \includegraphics[width=\linewidth]{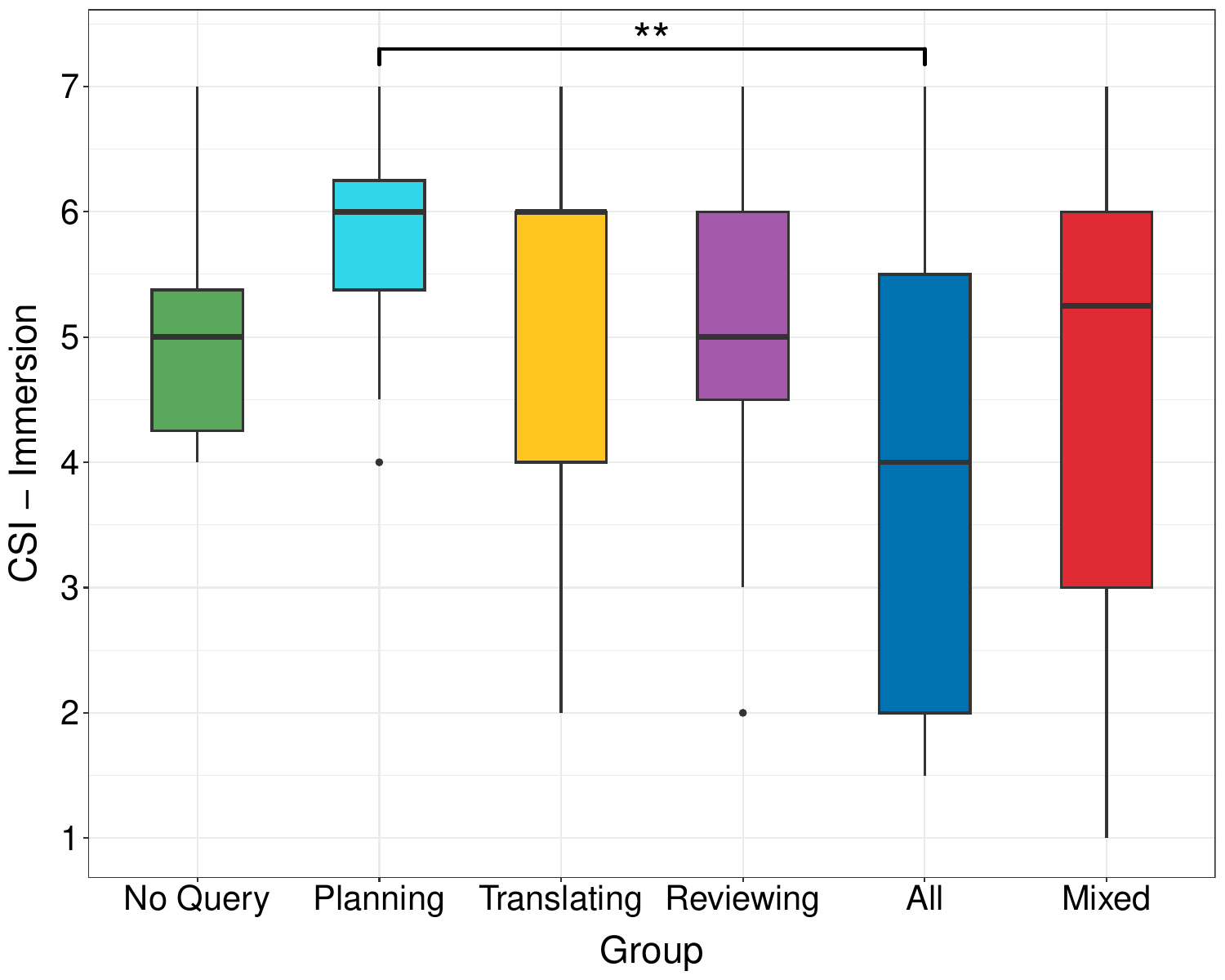}
    \caption{CSI-Immersion across Group}
    \Description{Box plots showing Creativity Support Index subscale Immersion scores on zero to seven scale. Varying clusters for the groupings, with Planning and All showing significant differences at 6 and 4 respectively.}
    \label{fig:csi_imm}
    \end{subfigure}
    \caption{Box Plots for CSI-Exlploration and CSI-Immersion with Post Hoc analysis (* indicates $p < 0.05$)}
    \Description{Two box plots showing the significance in two of the Creativity Support Index subscales. The leftmost graph shows CSI-Exploration while the rightmost shows CSI-Immersion distribution}
    \label{fig:csi_box}
\end{figure*}

For the CSI subscales, we found significant effects in Exploration and Immersion.
Through Kruskal-Wallis and post hoc analysis for CSI-Exploration (shown in Figure \ref{fig:csi_exp}), we found a $H(5) = 11.110$, $p = .0490$ with a significant pair in Groups N and M. This indicates that Group M is able to generate and track their ideas better than the users in Group N.
For the CSI-Immersion subscale, we find a $H(5)=12.463$ and $p=.0300$. Post-hoc analysis showed that Groups P and A were statistically different (shown in Figure \ref{fig:csi_imm}). This shows that the users in the Planning group were able to be absorbed into the activity more than the users in the All group.

\subsection{Dataset}
\label{sec:dataset}
The data used in this study is made publicly available through a data repository. Those interested in the dataset can find it at \url{\datalink}.
    
\section{Discussion}
In summary, our study provides rich insights into how students use ChatGPT for writing assignments and how their usage relates to their backgrounds, essay characteristics, and perceptions. 
 Building on \citeauthor{flower_cognitive_1981}' Cognitive Process Theory of Writing, we developed a taxonomy categorizing how participants' queries relate to different processes of the writing process; queries were grouped into Planning, Translating, Reviewing, and All categories. We further identified how factors such as writing self-efficacy, technology acceptance, and demographics influence the frequency of ChatGPT use. A clustering analysis revealed common usage patterns based on the distribution of query types students asked, with perceptions of both their essays and ChatGPT varying accordingly. For example, students who used ChatGPT to generate entire essays reported lower perceived ownership compared to those who engaged with it primarily for Planning and Reviewing.

\subsection{From Planning to Vibe Writing: Educational Implications of ChatGPT Use}

Our findings raise important concerns for writing instructors. 
 When given the opportunity to use ChatGPT, a substantial number of participants engaged in vibe writing, delegating most of the writing process to it. Even when students did not fully outsource their writing, many delegated critical stages such as ideation and opinion formation. This delegation effectively bypasses the processes through which students would otherwise critically engage with the topic, conduct research, translate ideas into a coherent argument, and self-evaluate their work for improvement.

Such behavior aligns with broader concerns that GenAI reduces the perceived effort of critical thinking and fosters overreliance on AI, potentially diminishing independent problem-solving skills among knowledge workers~\cite{10.1145/3706598.3713778}. Prior studies have also shown that GenAI use can lower cognitive load and even reduce brain activity in ways that may impair cognitive ability~\cite{STADLER2024108386, kosmyna_your_nodate}. Another thread of research showed that lack of critical engagement can result in compliance with GenAI's suggestions~\cite{jakesch2023co, bhat_interacting_2023}. Our study sheds light on previously hidden student practices and provides a taxonomy to guide future research into the mechanisms by which GenAI use influences learning in the context of writing.

The taxonomy we developed for RQ1 is consistent with the findings of \citeauthor{black2025university} which distinguishe two broad uses of ChatGPT: from lower-order writing tasks (e.g. proofreading, editing) to higher-order tasks (e.g. understanding complex topics, locating evidence)\cite{black2025university}. Our categories naturally map onto this framework: Planning aligns with higher‑order work; Reviewing aligns with lower‑order work; and Translating and All correspond to ghostwriting tasks. We extend this prior work by (1) specifying detailed codes within each category of the writing process and (2) contributing empirical interaction data (student queries and model responses) from a specific essay task, allowing deeper analysis and addressing the limitations of self‑reported descriptions of ChatGPT use in previous research~\cite{black2025university}.

A natural next step is to investigate instructors’ perceptions of the types of queries students make to ChatGPT, using our taxonomy as a reference, and to examine how instructors anticipate the potential learning impacts within their courses. Our study helps address limitations in existing research that rely on self-reported surveys from teachers, which are constrained by their limited understanding of how students actually use ChatGPT~\cite{barrett_not_2023, bower_how_2024}. In addition,  apart from readability scores, we did not analyze the content or quality of the essay, which could provide further insight into how ChatGPT affects not only student learning, but also assessment practices. For example, students who copied and pasted the prompt produced nearly identical essays, which instructors could readily identify as plagiarism. In contrast, students who engaged in vibe writing—actively shaping the essay at a high level and iteratively guiding GenAI—produced work that appeared original and was far more difficult for instructors to recognize as AI-assisted, as demonstrated in recent work~\cite{zeng2024detecting}. This future work can highlight a vulnerability that instructors may face in evaluating student writing in practice.

Another key takeaway from the study results is that the queries that a student asks GenAI alone cannot account for the learning impact that they may have for writing. Some participants used ChatGPT as if it were an expert to ask opinions about, and directly used the text that it generated, while others asked it to generate the entire essay, only to use it as a reference for their own independent writing. This complexity contributes to the challenge of understanding how the usage of LLM would impact a student's learning and regulate their usage per query type when banning is not an option. From our study, at least, the interaction trace that helps instructors have a better understanding will be responses that ChatGPT generates, and how the students consume it, which can manifest through following interaction (e.g., paste events, pause in GPT responses, etc).



\subsection{Student Personas of ChatGPT Use: Insights for Writing Instruction}

The identification of distinct yet recurring patterns in how students use ChatGPT is another contribution of this paper. Through our qualitative analysis of queries, we classified participants into six clusters of writers, suggesting that while individual use varied, their behaviors could be meaningfully grouped. This classification is further supported by the transition matrix and state diagram, which illustrate four critical usage paths (\autoref{fig:state}). Building on prior work that emphasized the challenge of classifying users~\cite{gero_social_2023, 10.1145/3635636.3656201, fitzsimmons_pressure_2025}, we identify six personas and quantify their prevalence within the participant pool, thereby characterizing common patterns of ChatGPT use in writing. 

Beyond Group A, the largest cluster, we identified two smaller groups of users who relied on ChatGPT exclusively for distinct purposes: generating ideas (Group P) or refining their final essay (Group R). We also observed a small group of non-users (Group N) who wrote their essays entirely on their own, potentially reflecting resistance to ChatGPT in practice, even when it was explicitly allowed.  This is also shown through the total number of ChatGPT words in the final essay. Those in Group P tended to have a small number of ChatGPT words, being significantly different than our All or Mixed groupings. This further suggests Group P as using ChatGPT as an ideation partner or search engine rather than a writer. In contrast, Group M and A exhibited widely different behaviors; the group was divisive, with some participants using ChatGPT-generated words entirely for their essay, while others asked for advice and wrote their essays entirely on their own. While prior work has documented patterns such as ideation and proofreading~\cite{ammari2025students, black2025university}, our study extends these findings by showing that students cluster into groups defined by consistent query types—including those who avoid the tool altogether. Previous research has linked hesitation to use ChatGPT to ethical and integrity concerns~\cite{chan_ai_2023}. Investigating why some students limit their use to a single function, or abstain entirely, could provide valuable insight into how they perceive the role of GenAI in the writing process.

One notable pattern we observed was the limited use of Translating, with Group T being smaller than any other group. This contrasts with prior work showing that professional creative writers valued translating assistance from GenAI to overcome writer’s block~\cite{gero_social_2023}. A likely explanation lies in motivation: creative writers strive for originality, whereas students in our study had little incentive to produce original essays, especially given the context of a voluntary online study. In contrast, the predominance of the Planning group in our data suggests that a greater hurdle for a larger number of students was deciding what to write, rather than knowing what to write but struggling to begin. This finding points to a gap in critical engagement with the assignment topic, suggesting that students may outsource the most cognitively demanding stage of writing to ChatGPT. Such reliance raises concerns for instructors whose goal is to help students critically examine specific topics, such as professionalism or ethics in computing.

 We further examined how ChatGPT influenced the final essay through readability scores, finding that users who had ChatGPT write text produced higher readability scores. These findings align with other studies where ChatGPT produced text with high complexity scores for technical writing~\cite{marulli_understanding_2024}, but ChatGPT received lower scores in Flesch-Kincaid Grade Level and Dale-Chall when used for creative writing~\cite{marulli_understanding_2024, romoff_role_2025}. We attribute our higher scores to the argumentative nature of the essay prompt.

\subsection{Student Backgrounds Shape How ChatGPT Is Used in Writing}


In RQ2, we investigated the relationship between students’ backgrounds and their use of ChatGPT. First, we found that lower writing self-efficacy was linked to more frequent querying overall, particularly for Translating and Reviewing. By contrast, Planning and All did not show the same pattern, suggesting that reliance on ChatGPT for tasks with greater implications for critical engagement may stem less from confidence in writing and more from external factors such as motivation or time pressure.

TAM, which captures students’ acceptance of ChatGPT, showed mixed results. Perceived usefulness (TAM PU) was positively associated with All queries, which may suggest that students who trusted the quality of ChatGPT’s output were more likely to generate entire essays with it. By contrast, perceived ease of use (TAM PEOU) was negatively associated with All queries; one possible explanation is that students who found ChatGPT easy to use may have had the skills to control it in more targeted ways, reducing the need to generate full essays. Similarly, the negative association between TAM PU and Reviewing queries may indicate a limited understanding of ChatGPT’s capabilities, leading students to see it merely as a proofreading tool. This result adds nuances to existing findings where they found that technology acceptance was correlated with the frequency of ChatGPT use~\cite{10.1145/3745238.3745434}.

We also found that gender, age, and race were significant predictors of the Total number of queries submitted. Participants who identified as White or Male submitted significantly more queries, particularly in the Reviewing category. Prior research suggests that women typically report higher self-efficacy in writing and reading~\cite{Pajares2006}, which may help explain why men were more inclined to turn to AI for writing support, as men may have less confidence in their writing skills. This interpretation aligns with our own finding of a negative association between Reviewing queries and writing self-efficacy. Other studies have shown that men hold more positive attitudes toward AI — using tools like ChatGPT to validate their work~\cite{grassini_gender}—and report greater awareness and perceived knowledge about AI~\cite{cachero_gender_2025}. Interestingly, these perspectives do not align with our result, where TAM PU was negatively associated with Reviewing queries. Age also played a role: a recent study found that younger generations feel pressured to use GenAI in contexts such as university applications~\cite{fitzsimmons_pressure_2025, madden2024dawn}, pointing to distinct adoption patterns. These findings provide a more nuanced understanding of how demographic factors shape GenAI use in writing by informing which type of usage is relevant.

It is worth mentioning that Planning queries did not have any significant predictors. The result may suggest that idea generation is a universal hurdle in writing, one less dependent on confidence, tool perceptions, or demographics, and more influenced by situational factors such as topic familiarity or their intrinsic/extrinsic motivation on a topic or assignment.
These interpretations indicate important directions for future research, examining how students’ writing background, perceptions of ChatGPT, and demographic characteristics influence both how often they use it and how they engage with it in the writing process.

\subsection{How ChatGPT Usage Impacts Ownership and Creative Engagement in Writing}

We examined how students’ writing experiences varied depending on their use of ChatGPT, focusing on Perceived Ownership (PO) and the Creativity Support Index (CSI). We found that PO differed significantly across three pairs of groups: N and A, P and A, and R and A. Unsurprisingly, those who generated an entire essay with ChatGPT felt less ownership than students who engaged with the writing process more directly. \citeauthor{joshi_writing_2025} reported that providing more detailed, content-rich queries leads to higher perceived ownership~\cite{joshi_writing_2025}. This helps explain why students who simply copied and pasted the writing prompt into ChatGPT—investing minimal effort—reported lower ownership. 

However, a more surprising—and perhaps concerning—finding is that the perceived ownership reported by all other groups who used ChatGPT (P, T, R, M) was comparable to, i.e., not significantly different from, that of Group N, who never used it at all. Despite missed learning opportunities, these students may still have felt that they wrote the essay. This suggests that the selective use of ChatGPT does not reduce students’ sense of ownership, even though it may deprive them of the chance to critically engage with the topic, evaluate their work, and revise their writing carefully.

Lastly, students’ perceptions of how ChatGPT supported their creative practice revealed subtle differences in two dimensions of the Creativity Support Index: Exploration and Immersion. For Exploration, Group M reported higher support than Group N. Unlike Group A, who primarily relied on ChatGPT to generate text, Group M used it more broadly—for idea generation, revision, and essay construction. They also submitted the largest number of queries overall (\autoref{fig:groupinquiry}), distributed relatively evenly across Planning, Reviewing, and All (\autoref{fig:cluster_dist}). A recent experiment found that intelligent features designed to support ideation made writing more engaging, as measured by increased time spent in the ideation process, though not necessarily in the reviewing process~\cite{goldi_support_2024}. This aligns with our finding that Group M may have engaged in exploring and comparing multiple ideas at a high level, rather than simply delegating their writing or focusing narrowly on expression.

For Immersion, Group P reported higher levels than Group A. Group A had the highest proportion of GPT-pasted words, indicating that their writing experience was shaped largely by text generation. In contrast, Group P concentrated on refining and engaging with the writing process itself, which may have fostered deeper immersion. In some respects, Group P appeared even more focused on writing than Group N, who had to both generate ideas and translate them into words. The ability to concentrate on shaping text may thus enhance immersion, potentially making the writing experience more engaging. 

 Prior work has already documented how GenAI can scaffold idea generation and support creative expression in various contexts~\cite{gero_social_2023, chakrabarty_art_2024, goldi_support_2024, lee_coauthor_2022, 10.1145/3635636.3656201}. Our results extend this literature by showing that not all GenAI use is equal: the kinds of queries students make correspond to meaningful differences (or the absence thereof) in specific subcomponents of creative engagement.
These results highlight how GenAI can subtly influence creative processes, providing creative professionals and toolmakers with a scaffold for certain types of queries in creative practice to emphasize a particular aspect of their experience (e.g., immersion or exploration).

\subsection{ Enabling Fine-Grained Analysis of AI-Assisted Writing}

 Our dataset offers value to future research in AI-assisted writing. Researchers in HCI, education, and natural language processing can use this dataset to explore the evolving nature of authorship in AI-assisted contexts. We provide a dataset offering keystroke-level trace data capturing the complete writing process, information not in other datasets~\cite{lee_coauthor_2022, han_recipe_2023, han_recipe4u_nodate, liu_detectability_2024, chatterji_how_2025, laskar_systematic_2023}. This granularity enables researchers to view how students engaged with ChatGPT over time — whether they gradually incorporated ideas, ignored them, or pasted them into the editor. Additionally, our dataset extends the existing literature by providing work from native English speakers, thereby broadening the scope for comparative research across various educational settings. This process-level data also stimulates research into temporal or causal analysis, work previously difficult or impossible. Researchers can investigate how the timing of AI consultation shapes their writing. The data also reveals critical decision points in a student's writing, such as where students choose to accept, modify, or reject AI suggestions. Furthermore, the data can be used to develop tools for educators, such as real-time dashboards that flag students who exhibit over-reliance on AI, or automated systems that provide targeted interventions based on detected usage patterns. By making this dataset publicly available, we anticipate supporting the research community in its efforts to gain a deeper, evidence-based understanding of how generative AI is reshaping writing education.


\section{Limitations and Future Work}


One potential limitation of this work is its ecological validity. Our study was conducted in a low-stakes environment where there were no negative consequences for using ChatGPT in ways that might otherwise be considered cheating. In real classrooms, writing assignments occur under a range of conditions (e.g., timed in-class exercises, untimed homework), whereas our study used only a soft 30-minute constraint. Deploying this kind of system in authentic classroom settings also raises ethical challenges, particularly around assessment, and students’ fear of being monitored may discourage authentic use. In addition, we believe that our study captures a common scenario in which students aim to invest just enough effort to produce an essay that they feel is ``good enough'' to submit.

 It is important to note that our study setting may have implicitly encouraged participants’ use of ChatGPT for various reasons: the task was framed as ``a study investigating essay writing and ChatGPT,” the ChatGPT window was visible alongside the editor, and participants may have been motivated to finish quickly given the lack of explicit rewards. We acknowledge that such factors could have amplified certain behaviors (e.g., Vibe Writing) observed in the study. However, we also anticipate that this encouragement resembles real situations in which students are tempted to use GenAI—for example, when facing a difficult topic, feeling unmotivated, or having limited time to complete an assignment. Even if some participants used ChatGPT more frequently than they would in authentic classroom settings, the range of behaviors we observed was broad, and most participants still completed the essay without ChatGPT generating major sections. Thus, while institutional or classroom policies may influence the frequency of GenAI use that we found in this paper, we believe that the behavioral patterns we identified (e.g., the taxonomy and the mode of vibe writing) remain meaningful.
 Similarly, because each student was clustered into one of six archetypes, the size of each group may vary in other contexts, but the relationships we identified—such as differences in essay characteristics or perceptions of ownership and creative engagement—should remain largely consistent. It remains an important area for future work to examine how real-world context and class policies shape students’ GenAI behaviors, including cases where students may conceal or misrepresent their usage. 

To address this limitation, we plan to deploy a classroom writing environment that integrates an in-house version of ChatGPT, where use is guided and regulated by instructor-defined GenAI policies. In this accountable AI platform, (1) students’ queries are made transparent to instructors, (2) instructors provide explicit guidance on acceptable use, and (3) the in-house ChatGPT adapts its responses according to the instructor’s specifications. Building on the taxonomy developed in this study, the system will classify student queries and either scaffold learning or decline to answer, depending on the policy. To ensure practical enforceability, the platform can restrict external GenAI use—for example, by disabling copy–paste from outside sources or embedding the model within a lockdown browser. Such design choices make the system realistic for classroom deployment while also enabling instructors to experiment with different policies and generate empirical evidence on how instructional contexts shape both student learning and ethical engagement with AI.

 One missed opportunity in this work, which motivates our immediate future work, is understanding how their ChatGPT usage pattern impacts the quality of the essay; we did not have human evaluations of essay quality and originality. Human evaluators (e.g., instructors) can provide greater insights regarding argument coherence, creativity, and critical thinking, dimensions not captured through readability scores, and relate them to how they used ChatGPT.In our future study, we will provide the final essay to instructors without ChatGPT histories to grade the essay. Then we can open students' AI history and study how instructors change their opinions, using this information to understand the learning impacts they can anticipate per query type or per student archetype. This will provide further insight into the impact of AI on student writing and inform policy regarding the use of AI in writing instruction.

\begin{acks}
We thank the reviewers for their constructive and insightful feedback, which helped improve this work, and we are grateful to all participants for their time and contributions. This research was supported in part by funding from the Center for Human Computer Interaction (CHCI) Planning Grant at Virginia Tech, and the 4VA grant. The views expressed in this material are those of the authors and do not necessarily reflect the views of the funding agencies.
\end{acks}

\bibliographystyle{ACM-Reference-Format}
\bibliography{gptwriting}


\appendix
\section{Appendix}
\subsection{Pre-study Survey}
\label{app:pre_survey}
\textbf{What gender do you identify as?}
\begin{itemize}
    \item Male
    \item Female
    \item Non-Binary
    \item Prefer not to disclose
    \item Other: \underline{\hspace{5cm}}
\end{itemize}

\textbf{How old are you?}
\begin{itemize}
    \item 18-24
    \item 25-34
    \item 35-44
    \item 45-55
    \item 55-64
    \item 65+
\end{itemize}

\textbf{What race/ethnicity describes you?}
\begin{itemize}
    \item American Indian or Alaskan Native
    \item Asian/Pacific Islander
    \item Black or African American
    \item Hispanic
    \item White/Caucasian
    \item Other: \underline{\hspace{5cm}}
\end{itemize}

\textbf{[Questions adapted from Self Efficacy in Writing~\cite{bruning2013examining}]}
\textit{Please answer the following statements with a 7-point Likert scale:}
\begin{enumerate}
    \item I can think of many ideas for my writing
    \item I can transform my ideas into written text
    \item I can think of many words to describe my ideas
    \item I can come up with many new ideas
    \item I know exactly how to organize my ideas into my writing
    \item I can spell my words correctly
    \item I can write complete sentences
    \item I can punctuate correctly, i.e., put punctuation marks such as full stops and commas in my sentences
    \item I can write grammatically correct sentences
    \item I can begin my paragraphs in the right spots
    \item I can focus on my writing for at least one hour
    \item I can ignore distractions while I’m writing
    \item I can start writing assignments quickly
    \item I can control my frustration while I’m writing
    \item I can think of my writing goals before I write
    \item I can keep writing even when it’s difficult
\end{enumerate}

\textbf{How often do you use ChatGPT for writing tasks?}
\begin{itemize}
    \item Never
    \item Once in a while
    \item About half the time
    \item Most of the time
    \item Always
\end{itemize}

\textbf{[Questions adapted from TAM~\cite{davis_perceived_1989}]}
\textit{For the following, please answer based on your usage of ChatGPT (7-point Likert scale):}
\begin{enumerate}
    \item Using ChatGPT would enable me to accomplish writing tasks more quickly
    \item Using ChatGPT increases my performance in writing tasks
    \item Using ChatGPT increases my productivity in writing tasks
    \item Using ChatGPT would enhance my effectiveness in writing tasks
    \item ChatGPT makes writing tasks easier for me
    \item I have found ChatGPT useful in writing tasks
    \item Learning to use ChatGPT would be easy for me
    \item I find it easy to get ChatGPT to do what I want it to do
    \item My interactions with ChatGPT are clear and understandable
    \item I find ChatGPT flexible to interact with
    \item It would be easy for me to become skillful at using ChatGPT
    \item I find ChatGPT easy to use
\end{enumerate}

\textit{For the following, please answer based on your usage of ChatGPT (7-point Likert scale):}
\begin{enumerate}
    \item I leverage the advanced features of ChatGPT to achieve my goals more efficiently than other students
    \item I’m often interested in trying new features
    \item I maximize the capabilities of ChatGPT
\end{enumerate}

\textbf{What is the highest degree or level of school you have completed or are currently pursuing?}
\begin{itemize}
    \item No schooling completed
    \item Some high school, no diploma
    \item High school graduate, diploma or equivalent (e.g., GED)
    \item Some college credit, no degree
    \item Trade/technical/vocational training
    \item Associate degree
    \item Bachelor’s degree
    \item Master’s degree
    \item Professional degree
    \item Doctorate degree
\end{itemize}

\textbf{Have you completed the degree specified above?}
\begin{itemize}
    \item Yes
    \item No
\end{itemize}

\textbf{What is your current major?} \underline{\hspace{5cm}}

\subsection{Writing Prompt}
\label{app:writing}

We will describe an issue and provide three different perspectives on the issue. You are asked to read and consider the issue and perspectives, state your own perspective on the issue, and analyze the relationship between your perspective and at least one other perspective on the issue.

\textbf{Issue:
}

Automation is generally seen as a sign of progress, but what is lost when we replace humans with machines?

\textbf{Intelligent Machines}

Many of the goods and services we depend on daily are now supplied by intelligent, automated machines rather than human beings. Robots build cars and other goods on assembly lines, where once there were human workers. Many of our phone conversations are now conducted not with people but with sophisticated technologies. We can now buy goods at a variety of stores without the help of a human cashier. Automation is generally seen as a sign of progress, but what is lost when we replace humans with machines? Given the accelerating variety and prevalence of intelligent machines, it is worth examining the implications and meaning of their presence in our lives.

\textbf{Perspective One (Dystopian view)}

What we lose with the replacement of people by machines is some part of our own humanity. Even our mundane daily encounters no longer require from us basic courtesy, respect, and tolerance for other people.

\textbf{Perspective Two (Utilitarian view)
}
Machines are good at low-skill, repetitive jobs, and at high-speed, extremely precise jobs. In both cases they work better than humans. This efficiency leads to a more prosperous and progressive world for everyone.

\textbf{Perspective Three (Progressive view)
}
Intelligent machines challenge our long-standing ideas about what humans are or can be. This is good because it pushes both humans and machines toward new, unimagined possibilities.

\subsection{Exit Survey}
\label{post_survey}
    Thank you for participating in our study. Please answer the following questions as part of our exit survey.
    
    \textbf{[Questions adapted from PO~\cite{avey_psychological_2009, chantal_psychological_2012}] For the following questions, please answer based on your perceived ownership of the essay:}
    \textit{7-point Likert Scale}
    
    \begin{enumerate}
        \item I feel that this is my essay
        \item I feel that this essay belongs to me
        \item I feel a high degree of ownership towards this essay
        \item I feel the need to protect my ideas from being used by others.
        \item I feel that this essays success is my success
        \item I feel this essay was written by me
        \item I feel the need to protect the ideas written in the essay
        \item I do not feel like anyone else wrote this essay.
    \end{enumerate}

    \textbf{[Questions adapted from CSI~\cite{cherry_quantifying_2014}] For the following questions, please answer based on your usage of ChatGPT:}
    \textit{7-point Likert Scale}
    
    \begin{enumerate}
        \item I feel like ChatGPT helped me in the creation process of my writing
        \item I feel like ChatGPT helped me with proofreading my essay
        \item I feel like ChatGPT made my essay better
        \item I liked using ChatGPT as an assistant during my essay writing
        \item My writing would have been better without ChatGPT assistance
    \end{enumerate}
    
    Thank you for completing our survey. 

\subsection{Clusters}
\begin{figure*}[!htbp]
    \centering
    \begin{subfigure}{0.49\textwidth} 
        \centering
        \includegraphics[width=\linewidth]{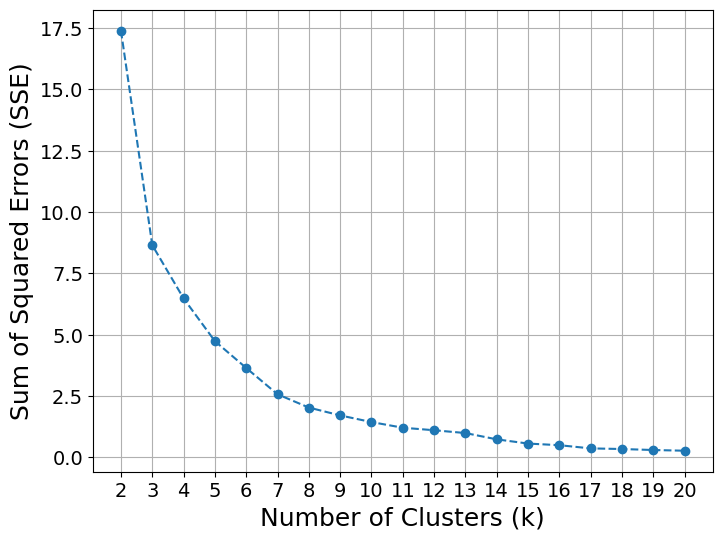}
        \caption{Sum of squared errors (SSE) per the number of clusters}
        \label{fig:SSE}
        \Description{Line Graph showing the SSE of the clusters used to determine the optimal cluster count. Elbow is at 6 clusters indicating proper cluster choice.}
    \end{subfigure}
    \hfill 
    \begin{subfigure}{0.49\textwidth}
         \centering
        \includegraphics[width=\linewidth]{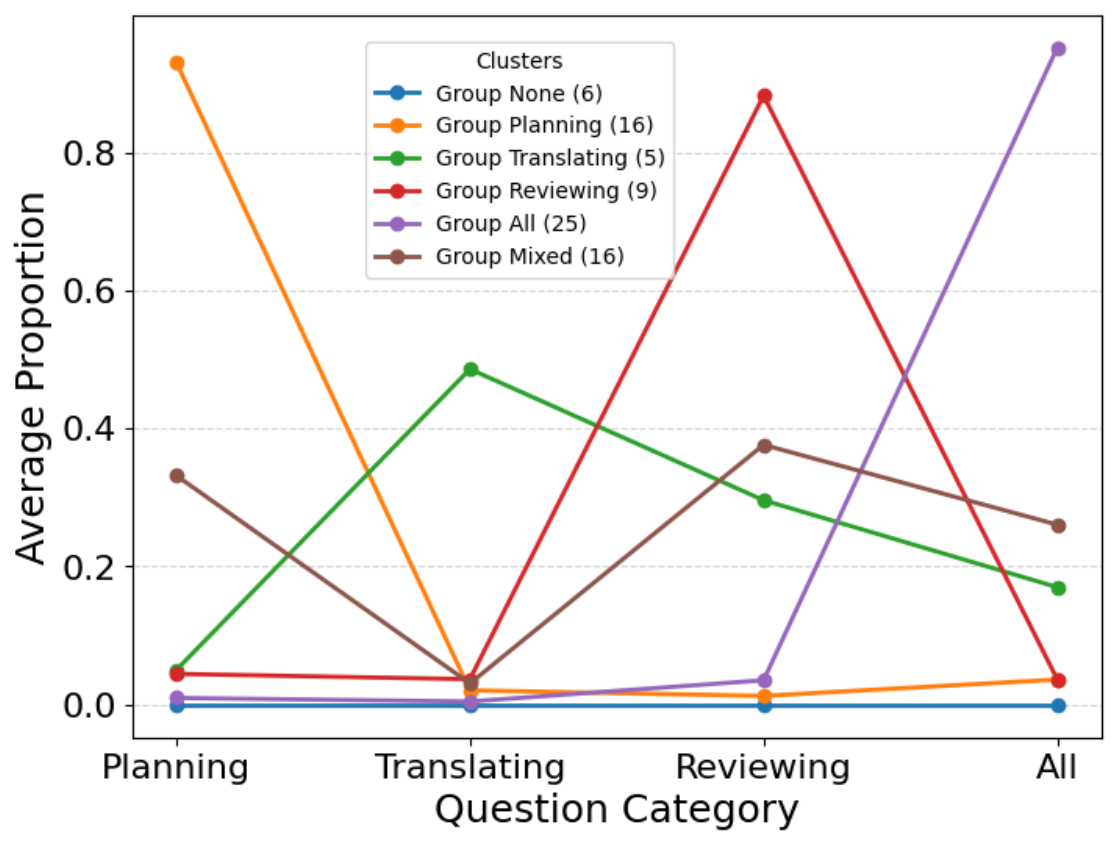}
        \caption{Distribution of our clusters}
        \Description{The proportion of each question code across the different clusters. The legend shows the count of each cluster.}
        \label{fig:cluster_prop}
    \end{subfigure}
    \caption{K-means clustering SSE and Cluster information}
    \Description{Details about the clusters and clustering method used for qualitative analysis. The leftmost graph shows the Sum of Squared Errors and the rightmost shows the distribution of our clusters}
    \label{fig:k_means}
\end{figure*}

\end{document}